\documentclass[pre,11pt,preprint,superscriptaddress, amsmath, amssymb, floatfix]{revtex4-1} 
\usepackage{graphicx}  
\usepackage{graphics}
\usepackage{dcolumn}
\usepackage{textcase}   
\usepackage{bm}        
\usepackage{amssymb}   
\usepackage{color}
\usepackage{multirow}
\usepackage{amsmath}
\usepackage{threeparttable}
\usepackage[font=scriptsize]{caption}
\usepackage{url}
\usepackage{makecell}
\usepackage{diagbox}
\usepackage{algorithm}
\usepackage{algorithmic}
\usepackage{booktabs}
\usepackage{subfigure}
\usepackage{amsfonts}
\usepackage{rotating}
\usepackage{fontenc}
\usepackage{cancel}
\usepackage{hyperref}
\usepackage[normalem]{ulem}
\usepackage{mathtools}

\usepackage{titlesec}
\titleformat{\section}[block]{\large\bfseries}{\thesection}{1em}{}
\titlespacing*{\section}{0pt}{*1}{*1}

\hypersetup{CJKbookmarks,
       bookmarksnumbered,
              colorlinks,
               linkcolor=black,
               citecolor=black,
              plainpages=false,
            pdfstartview=FitH}

\definecolor{darkred}{RGB}{139,0,0}
\definecolor{chartreuse}{RGB}{127,255,0}
\definecolor{goldenrod}{RGB}{218,165,32}
\definecolor{gray}{RGB}{127,127,127}

\definecolor{Magenta}{RGB}{255, 0,255}
\definecolor{Orange}{RGB}{255,165, 0}
\definecolor{Gray}{RGB}{127,127,127}

\begin{document}

\title{Uncovering multi-order popularity and similarity mechanisms in link prediction by graphlet predictors}
\author{Yong-Jian He}
\affiliation{Department of Systems Science, Faculty of Arts and Sciences, Beijing Normal University, Zhuhai 519087, China}
\affiliation{International Academic Center of Complex Systems, Beijing Normal University, Zhuhai 519087, China}
\affiliation{School of Systems Science, Beijing Normal University, Beijing 100875, China}

\author{Yijun Ran}
\affiliation{Computational Communication Research Center, Beijing Normal University, Zhuhai 519087, China}

\author{Zengru Di}
\affiliation{Department of Systems Science, Faculty of Artsand Sciences, Beijing Normal University, Zhuhai 519087, China}
\affiliation{International Academic Center of Complex Systems, Beijing Normal University, Zhuhai 519087, China}

\author{Tao Zhou}
\email{zhutou@ustc.edu, xuxiaoke@foxmail.com}
\affiliation{CompleX Lab, University of Electronic Science and Technology of China, Chengdu 611731, China}
\affiliation{Big Data Research Center, University of Electronic Science and Technology of China, Chengdu\ 611731, China}

\author{Xiao-Ke Xu}
\email{zhutou@ustc.edu, xuxiaoke@foxmail.com}
\affiliation{Computational Communication Research Center, Beijing Normal University, Zhuhai 519087, China}
\altaffiliation{Both authors contributed equally to this research and are joint corresponding authors.}

\begin{abstract}
Link prediction has become a crucial problem in network science and has thus attracted increasing research interest. Popularity and similarity are two primary mechanisms in the link formation of real networks. However, the roles of these mechanisms in link prediction across various networks in domains remain poorly understood. Here, we use orbit degrees of graphlets to construct lower- and higher-order predictors and demonstrate that multi-order popularity and similarity indices can be efficiently represented in terms of orbit degrees. We further propose a supervised learning model that fuses multiple orbit-degree-based features and validate its enhanced performance. We observe that lower-order features, particularly lower-order similarity, serve as the primary role in the link prediction, while higher-order features tend to play a complementary role. Our research improves the accuracy and interpretability of link prediction and provides new perspectives and tools for analyzing complex networks.
\end{abstract}

\maketitle

\section*{\large\bfseries Introduction}
\hspace*{1em} Networks constitute a powerful tool for analyzing complex systems, such as social, biological, and technological systems, with network nodes and links representing elements and the interactions between elements, respectively \cite{albert2002statistical,newman2003structure}. Link prediction is a fundamental network-related problem, aiming to infer the existing likelihood of a link between a pair of nodes that are not connected by leveraging partial network structure information \cite{lu2011link,zhou2021progresses}. The predicted links comprise links that exist but have yet to be discovered (missing links) and links that do not currently exist but may appear in the future (future links) \cite{lu2011link,zhou2021progresses}.
Link prediction serves not only as a criterion for evaluating and comparing network evolution models but also contributes to a deeper understanding of the organizational principles governing complex networks \cite{wang2012evaluating,zhang2015measuring,valles2018consistencies,ghasemian2019evaluating}. Consequently, link prediction finds applications in various domains, such as aiding in the recommendation of friends in social networks \cite{aiello2012friendship} and products in e-commercial websites \cite{lu2012recommender}, and guiding biological experiments to significantly reduce unnecessary expenses \cite{clauset2008hierarchical,kovacs2019network}.

In network science, link prediction heuristics attempt to extract information from a network structure to quantify the probability of the existence of links between nodes \cite{lu2011link, ren2018structure, kumar2020link,ran2024maximum}. For example, the Preferential Attachment (PA) index represents a node's popularity by its degree and posits that node with higher popularity (i.e., larger degree) have a greater probability of attracting other nodes to form links \cite{barabasi1999emergence}. The Common Neighbors (CN) index, based on the homophily principle, suggests that the more common neighbors two nodes share, the higher their similarity, and thus the greater the probability to form a link between them \cite{liben2007link}. The PA and CN respectively represent two kinds of essential mechanisms in network link formation, namely the popularity mechanism and the similarity mechanism. The formation of links in the famous scale-free network model is primarily based on the popularity mechanism of PA representation \cite{barabasi1999emergence}. Recent research has shown that link formation in real technological, social, and biological networks is driven by an optimized balance between similarity and popularity mechanisms, rather than relying solely on either one of these two mechanisms \cite{papadopoulos2012popularity}.

The PA and CN methods utilize lower-order (only accounting for first-order neighbors) information for link prediction, representing lower-order popularity and similarity mechanisms, respectively. Recent studies have found that higher-order information is more capable of capturing missing links in networks \cite{liu2019link,liu2019link2}, especially in networks containing rich higher-order structures. Therefore, scholars have proposed numerous innovative higher-order popularity and similarity methods \cite{zhou2009predicting,lu2009similarity}, such as motif-based approaches \cite{milo2002network, zhang2013potential,xia2019survey, liu2019link,liu2019link2,wang2020model,li2021research,xia2021chief}. At the same time, advanced embedding-based \cite{perozzi2014deepwalk,grover2016node2vec,cao2019network,mara2020benchmarking} and deep learning \cite{wang2020link,zhang2018link,krenn2023forecasting, muscoloni2023stealing,menand2024link,ghasemian2020stacking} methods are shown to be efficient in solving link prediction tasks. Unfortunately, so far, learning-based methods have not been deeply integrated with mechanism-based methods, or at least the optimal way to combine them has not been found. Additionally, existing analyses do not inform us which one, among the many popularity and similarity mechanisms, plays the most significant role in link formation, nor do they tell us whether higher-order information is relevant and, if so, to what extent. Furthermore, they do not indicate whether these answers would differ for networks in different domains.

In this work, we propose a novel graphlet-based link prediction framework. We use the orbit degrees of graphlets to quantify the multi-order popularity and similarity features and then integrate these features into a machine learning model for link prediction. By measuring features' importance, we uncover the roles of multi-order popularity and similarity mechanisms in link prediction of different networks. First, we show that many well-known popularity-based and similarity-based methods can be effectively represented using orbit degrees. Second, extensive experiments on 550 real-world networks from six domains indicate that the proposed orbit-degree-based model considerably outperforms other models. Final, we conduct the feature analysis and observe that first-order similarity features play a primary role in link prediction of most networks. In contrast, the higher-order similarity and multi-order popularity features play a complementary role. Social networks are mainly dominated by the first-order similarity mechanism expressed by homophily, which corresponds to the orbit degree M2$ (\vcenter{\hbox{\includegraphics[width=2.8ex,height=2.8ex]{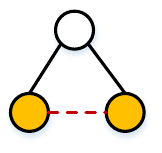}}})$. The orbit degree M3 $(\vcenter{\hbox{\includegraphics[width=2.8ex,height=2.8ex]{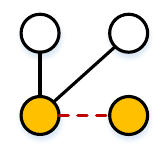}}})$ plays a leading role in link prediction within the technological, economic, and information networks. The presented framework is not only limited to link prediction but also serves as a versatile tool for addressing other critical issues in network science, such as identifying influential nodes and analyzing network spreading dynamics.

\section*{\large\bfseries  Results}

\begin{figure*}[htbp]
	\setlength{\abovecaptionskip}{0.cm}
	\setlength{\belowcaptionskip}{-0.cm}
	\centering
	\includegraphics[width=0.98\textwidth]{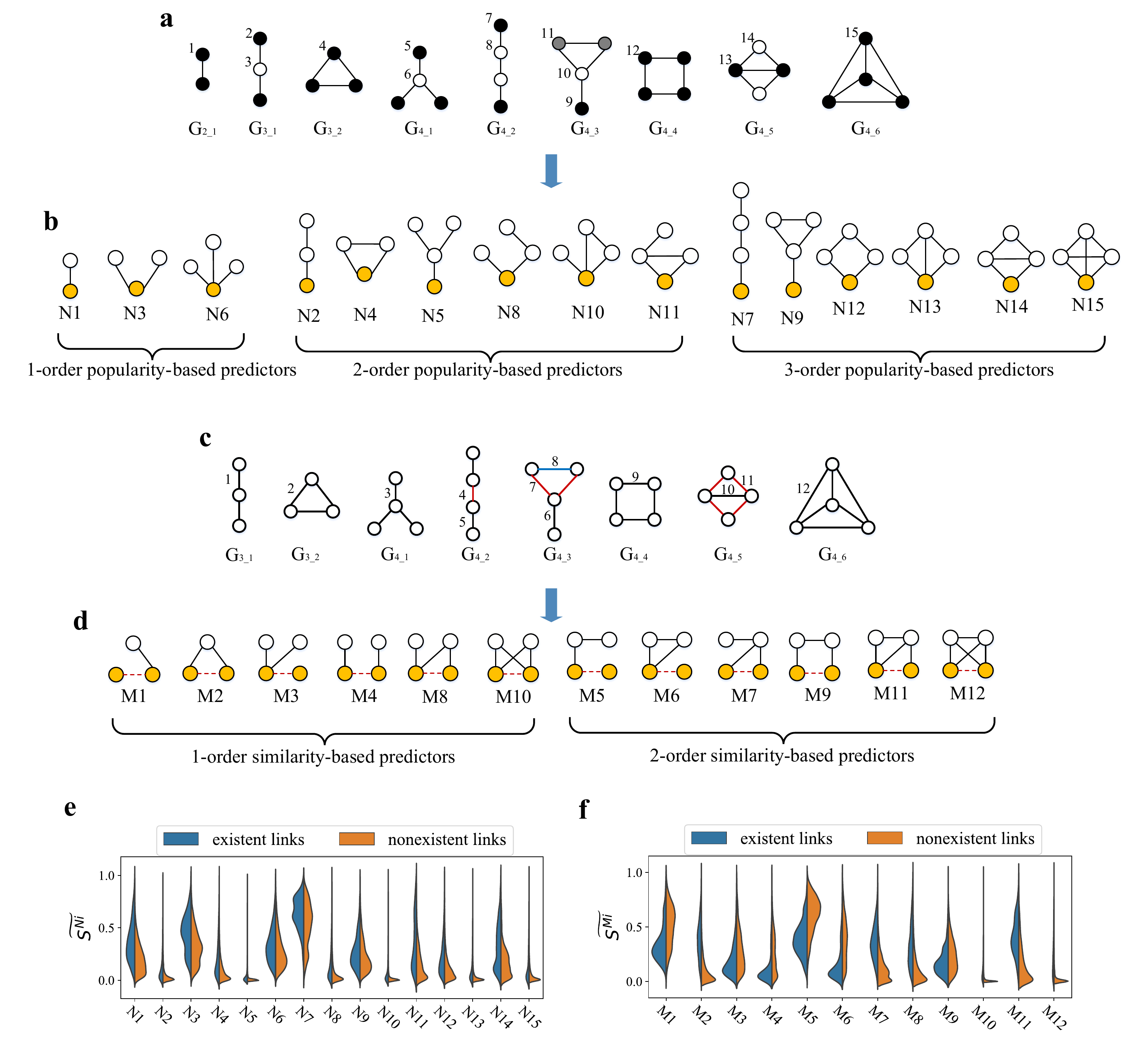}
	\caption{\textbf{Graphlet orbits, orbit-degree-based link predictors, and orbit degree distribution}.
		\textbf{a} Fifteen node orbits characterised by graphlets with 2\textendash4 nodes. In each graphlet, nodes with different gray levels belong to different node orbits.
		\textbf{b} Fifteen multi-order popularity-based predictors, with the yellow node in each node orbit representing one endpoint of the target link for prediction (called target node in the following text). The order is defined by the length of the longest path starting from the target node. In these features, we define first-order popularity-based predictors as lower-order features, and second- and third-order predictors as higher-order, with the same applied to similarity predictors.
		\textbf{c} Twelve edge orbits characterised by graphlets comprising 3\textendash4 nodes. In each graphlet, edges with different colors belong to different edge orbits.
		\textbf{d} Twelve multi-order similarity predictors, with the dotted red line in each edge orbit representing the target link for prediction. The order is defined by the length of the longest path starting from one endpoint without touching or passing through the other.
		\textbf{e} and \textbf{f} Distributions of normalized popularities $\widetilde{S^{Ni}_{xy}}=\frac{S^{Ni}_{\max}-S^{Ni}_{xy}}{S^{Ni}_{\max}-S^{Ni}_{\min}}$ and normalized similarities $\widetilde{S^{Mj}_{xy}}=\frac{S^{Mj}_{\max}-S^{Mj}_{xy}}{S^{Mj}_{\max}-S^{Mj}_{\min}}$ for existent and nonexistent links in a contact network of high school students, where $S^{Ni}_{\max}$ and $S^{Mj}_{\max}$ are maximum values, and $S^{Ni}_{\min}$ and $S^{Mj}_{\min}$ are minimum values over considered links for Eq. \ref{Equation_N} and Eq. \ref{Equation_M}, respectively. For most orbit degrees, the distributions for existent and nonexistent links are visually distinct.}
	\label{Fig1_Graphlets}
\end{figure*}

\hspace*{1em}{\bf Graphlet-based approach.}
Graphlets, defined as small, connected, nonisomorphic subgraphs within a larger network, constitute a critical tool for delineating and quantifying the local structural attributes of networks \cite{prvzulj2004modeling}.
The utilization of graphlets facilitates a detailed analysis of the topological configurations surrounding individual nodes and edges \cite{prvzulj2004modeling, prvzulj2007biological}. In general, nodes in a graphlet can belong to different automorphism orbits (see nodes with different gray levels in Fig. \ref{Fig1_Graphlets}a), and thus it is insufficient to say that a node belongs to a graphlet without specifying its orbit. This is the same in the consideration of edges (see the edges with different colors in Fig. \ref{Fig1_Graphlets}c).

Formally, an isomorphism $f(\cdot)$ from graph $X$ to graph $Y$ is a bijection from the nodes of $X$ to the nodes of $Y$, such that if $(x,y)$ is an edge of $X$, then $(f(x),f(y))$ is an edge of $Y$. An automorphism $g(\cdot)$ is a special isomorphism that maps a graph to itself. The set of all automorphisms of a graph $X$ forms the automorphism group of $X$, which is denoted by $Aut(X)$. For an arbitrary node $x$ in graph $X$, the automorphism orbit of $x$ in the automorphism group is
\begin{equation}
	Norb(x) = \{ x' \in V(X) | x'=g(x)\},
\end{equation}
where $V(X)$ is the set of nodes of graph $X$ and $g \in Aut(X)$. Similarly, the automorphism orbit of an edge $(x,y)$ in graph $X$ is
\begin{equation}
	Eorb((x,y)) = \{(x',y') \in E(X)|x'=g(x), y'=g(y)\},
\end{equation}
where $E(X)$ is the set of edges in graph $X$. In Fig. \ref{Fig1_Graphlets}, the same gray level and color are used, respectively, to indicate nodes and edges within the same orbit in a graphlet. Graphlets comprising two to four nodes collectively encapsulate fifteen distinctive node orbits (Fig. \ref{Fig1_Graphlets}a) and twelve distinctive edge orbits (Fig. \ref{Fig1_Graphlets}c). We only focus on graphlets comprising fewer than five nodes for two reasons. First, the calculation of orbit degrees associated with small graphlets has low computational complexity. Second, higher-order graphlets are typically composed of lower-order graphlets, and the abundance of higher-order graphlets in a graph depends on the prevalence of the constitutive lower-order graphlets \cite{vazquez2004topological}.

Studies employing subgraph models for link prediction have predominantly focused on motifs \cite{milo2002network, zhang2013potential,xia2019survey, liu2019link,liu2019link2,wang2020model,li2021research,xia2021chief,backstrom2014romantic,dong2017structural}, where the number of motifs that a targeted link participates in served as a quantitative index for link prediction. However, motifs do not account for the varied roles of different edges within them, thereby ignoring edge orbits. A few studies \cite{milenkovic2008uncovering, hulovatyy2014revealing,feng2020link,he2023predicting} have utilized graphlets in link prediction, where they estimated the existing likelihood of target links by constructing Graphlet Degree Vectors (GDV) for nodes \cite{prvzulj2007biological} or edges \cite{solava2012graphlet}, based on the number of graphlets to which a node or edge belongs. However, nodes or edges belonging to the same graphlet but to different orbits have different structural roles. Therefore, in this study we focus on the frequency of each orbit instead, and define the so-called node orbit degree and edge orbit degree. Then, we utilize node orbit degree to quantify multi-order popularity of a node, and employ edge orbit degree to characterize multi-order similarity between two nodes.

The node orbit degree is defined as the number of graphlets that a node touches at the specific node orbit of the graphlets \cite{prvzulj2007biological}.
Formally, given a network $G$, for a specific graphlet $M$, we assume there are $n$ node sets $\{Norb_1, Norb_2, \cdots, Norb_n\}$ corresponding to the $n$ different node orbits in $M$, and $l$ instances of graphlet $M$, denoted as $\{M^1,M^2, \cdots ,M^l\}$. The node set corresponding to the $i$-th node orbit in the $c$-th instance is then denoted as $Norb_i^c$. The node orbit degree of any node $x$ is
\begin{equation}
	NOD_i(x) = \sum\limits_{c = 1}^l {N_i^c(x)},
\end{equation}
where
\begin{equation}
	N_i^c(x) = \begin{cases} 1, & x \in Norb_i^c\\
		0 ,& otherwise \\
	\end{cases}.
\end{equation}

The edge orbit degree is defined as the number of graphlets that an edge touches at the specific edge orbit of the graphlets \cite{solava2012graphlet}. Similarly, if there are $m$ edge sets corresponding to the $m$ different edge orbits in $M$, we use $Eorb_j^c$ to denote the set of edges on the $j$-th edge orbit in the $c$-th instance, and then the edge orbit degree of $(x,y)$ reads
\begin{equation}
	EOD_j(x,y) = \sum\limits_{c = 1}^l {E_j^c(x,y)},
\end{equation}
where
\begin{equation}
	E_j^c(x,y) = \begin{cases} 1, & (x,y) \in Eorb_j^c\\
		0 ,& otherwise \\
	\end{cases}.
\end{equation}

The algorithm implementation and complexity analysis of the node and edge orbit degrees are shown in \textcolor{blue}{Supplementary Note 1}.

We derive a total of fifteen multi-order popularity-based predictors using node orbit degrees, as displayed in Fig. \ref{Fig1_Graphlets}b, where each yellow node represents one endpoints of the link to be predicted. The order of a predictor is defined as the length of the longest path starting from the yellow node. For example, for N1$(\vcenter{\hbox{\includegraphics[width=1.5ex,height=2.8ex]{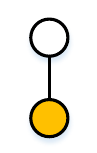}}})$, N3$(\vcenter{\hbox{\includegraphics[width=2.8ex,height=2.8ex]{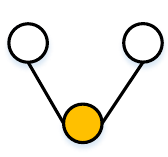}}})$ and N6$(\vcenter{\hbox{\includegraphics[width=2.8ex,height=2.8ex]{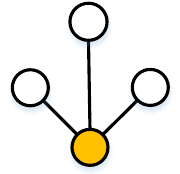}}})$, the longest path length starting from the yellow node is one, so they are all first-order popularity-based predictors. Analogously, N2$(\vcenter{\hbox{\includegraphics[width=2.8ex,height=2.8ex]{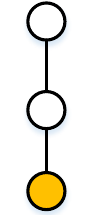}}})$ and N5$(\vcenter{\hbox{\includegraphics[width=2.8ex,height=2.8ex]{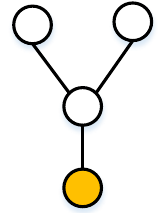}}})$ belong to the second-order predictors, and N7$(\vcenter{\hbox{\includegraphics[width=1.5ex,height=2.8ex]{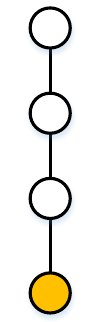}}})$ and N9$(\vcenter{\hbox{\includegraphics[width=2.8ex,height=2.8ex]{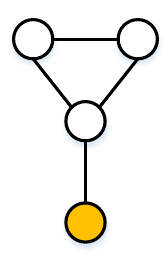}}})$ belong to the third-order predictors. For a target link, inspired by the PA index, we use the product of the node orbit degrees of its two endpoints to estimate its existing likelihood, say
\begin{equation}
	\begin{split}
		S^{Ni}_{xy}= {N_i(x)} \cdot {N_i(y)},
	\end{split}
	\label{Equation_N}
\end{equation}
where $N_i(x)$ is the node orbit degree of node $x$ for $Ni$. The corresponding fifteen predictors based on the node orbit degree are denoted as ${S^{N1}, S^{N2}, \cdots, S^{N15}}$.

In a similar manner, we construct twelve multi-order similarity-based predictors using edge orbit degrees as shown in Fig. \ref{Fig1_Graphlets}d, where the red dash links represent the target links (yellow node pairs) to be predicted. The order of a predictor is defined as the length of the longest path that starts from any one endpoint but does not touch the other endpoint. For example, the predictors M1$(\vcenter{\hbox{\includegraphics[width=2.8ex,height=2.8ex]{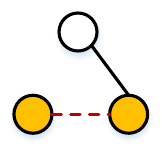}}})$ and M2$(\vcenter{\hbox{\includegraphics[width=2.8ex,height=2.8ex]{./motif/M2.pdf}}})$ are first-order similarity-based predictors. Analogously, M5$(\vcenter{\hbox{\includegraphics[width=2.8ex,height=2.8ex]{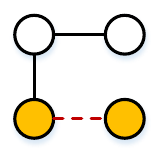}}})$ and M9$(\vcenter{\hbox{\includegraphics[width=2.8ex,height=2.8ex]{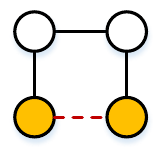}}})$ are the second-order predictors. Inspired by the CN index, we treat edge orbit degrees as similarity scores for the target node pair, say
\begin{equation}
	\begin{split}
		S^{Mj}_{xy}= M_j(x,y),
	\end{split}
	\label{Equation_M}
\end{equation}
where $M_j(x,y)$ is the edge orbit degree of edge $(x,y)$ for $Mj$. The corresponding twelve predictors base on the edge orbit degree are denoted as $S^{M1}, S^{M2}, \cdots, S^{M12}$.

It is important to note that, in the traditional link prediction approaches (e.g., CN), a higher link score always indicates a higher existing likelihood, while in the current method, an orbit degree may exert a negative role, namely a higher score may imply a lower likelihood for some orbit degrees. In our approach, each orbit degree is treated as a one-dimensional feature, and we draw the prediction based on a supervised binary classification model. Therefore, whether a feature contributes positively or negatively to the existing likelihood is secondary to its effectiveness in distinguishing between an existent link and a nonexistent one. As such, a feature that aids in this differentiation is considered to have contribution to the link prediction task, in despite of the possibility that a large feature value may correspond to a lower existing likelihood.

To provide a clear understanding of the functions of orbit degrees in link prediction, we compare the distributions of orbit-degree-based scores of node pairs for existent and nonexistent links in a contact network of high school students \cite{fournet2014contact} (Figs. \ref{Fig1_Graphlets}e and f). Scores based on the node and edge orbit degrees are calculated using Eqs. \ref{Equation_N} and \ref{Equation_M}, respectively. These distributions are visualized using violin plots. The key observation is that, in the majority of cases, the distributions for existent and nonexistent links are remarkably different.

{\bf Orbit-degree representations of existing multi-order popularity- and similarity-based methods.}
The proposed orbit-degree-based approach provides a systematic framework for multi-order popularity-based and similarity-based indices in link prediction, namely many well-known multi-order popularity-based and similarity-based methods can be regarded as specific cases of orbit degree methods. Here, we show in detail how certain well-known indices can be expressed by orbit degrees.

\begin{figure*}[htbp]
	\setlength{\abovecaptionskip}{0.cm}
	\setlength{\belowcaptionskip}{-0.cm}
	\centering
	\includegraphics[width=0.95\textwidth]{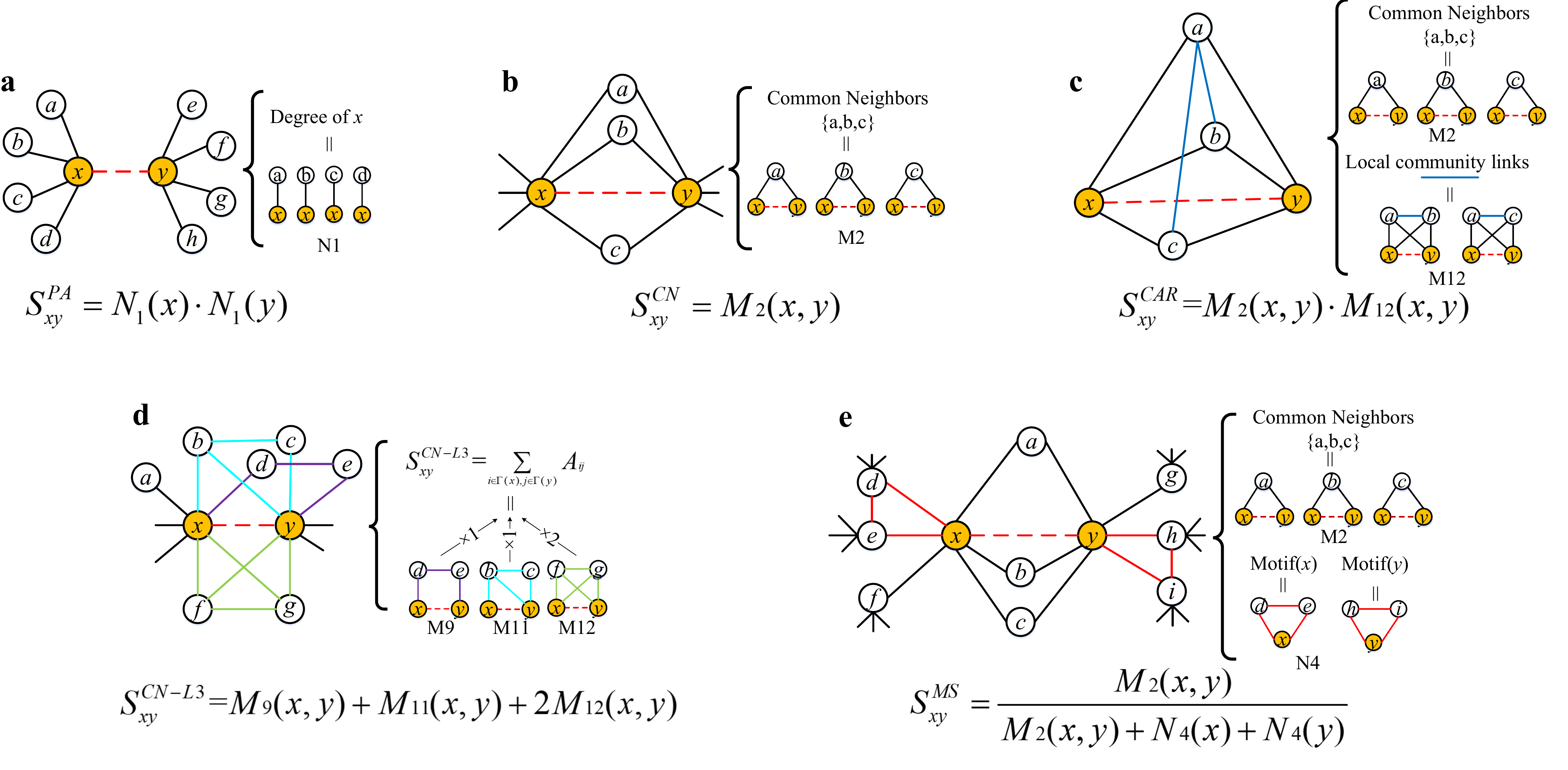}
	\caption{\textbf{Decomposing known popularity-based and similarity-based indices by orbit degrees.}
		\textbf{a} The degrees of nodes $x$ and $y$ are captured by orbit N1. Thus, the PA index of these nodes is simply the product of their numbers of N1 orbits.
		\textbf{b} The CN index indicates the number of common neighbors for nodes $x$ and $y$, which can be directly mirrored by M2.
		\textbf{c} The CAR index depends on M2, which reflects the common neighbors, and M12, which represents the local community links.
		\textbf{d} The CN-L3 index indicates the combined effects of edge orbits M9, M11, and M12 around nodes $x$ and $y$, with each $M_9({x,y})$ or $M_{11}({x,y})$ contributing one to the score $S_{xy}^{CN-L3}$ and each $M_{12}({x,y})$ contributing two.
		\textbf{e} The MS index measures the similarity between nodes $x$ and $y$ by looking at their common neighbors and triangles they form, which can be fully represented by M2 and N4.
	}
	\label{Fig2_Mechanisms}
\end{figure*}

The widely recognized Preferential Attachment (PA) index is a first-order popularity index \cite{barabasi1999emergence}, which is mathematically articulated as
\begin{equation}
	S_{xy}^{PA} = {k_x} \cdot {k_y},
	\label{PA}
\end{equation}
where $k_x$ and $k_y$ are the degrees of nodes $x$ and $y$, respectively. The degree of node $x$ is equivalent to the degree of node orbit $N_1(x)$ because N1$(\vcenter{\hbox{\includegraphics[width=1.5ex,height=2.8ex]{./motif/N1.pdf}}})$ represents the number of first-order neighbors of this node (Fig. \ref{Fig2_Mechanisms}a). Therefore, the PA index can be rewritten as
\begin{equation}
	S_{xy}^{PA} = {N_1(x)} \cdot {N_1(y)}.
	\label{Equation_PA}
\end{equation}

Common Neighbors (CN) index is the most known local similarity index in link prediction \cite{liben2007link}, which directly calculates the number of common neighbors of a pair of nodes. Denoting $\Gamma (x)$ as the neighbor set of any node $x$, then CN is defined as
\begin{equation}
	\begin{split}
		S_{xy}^{CN} = \left| {\Gamma (x) \cap \Gamma (y)} \right|.
	\end{split}
	\label{CN_c}
\end{equation}
Obviously, to count the edge orbit degree M2$(\vcenter{\hbox{\includegraphics[width=2.8ex,height=2.8ex]{./motif/M2.pdf}}})$ of two target nodes $x$ and $y$ is equivalent to counting their common neighbors(Fig. \ref{Fig2_Mechanisms}b). Therefore, the CN index can be reformulated as
\begin{equation}
	S_{xy}^{CN} = M_2(x,y).
	\label{CN}
\end{equation}

Employing the above approach, the Adamic-Adar (AA) index \cite{adamic2003friends} and Resource Allocation (RA) index \cite{zhou2009predicting}, which are well-known variants of the CN index, can also be formulated in terms of orbit degrees as
\begin{equation}
	S_{xy}^{AA} = \sum_{z \in \Gamma (x) \cap \Gamma (y)} \frac{1}{\log N_1(z)},
	\label{AA}
\end{equation}
and
\begin{equation}
	S_{xy}^{RA} = \sum_{z \in \Gamma (x) \cap \Gamma (y)} \frac{1}{N_1(z)}.
	\label{RA}
\end{equation}

Another representative variant of the CN is the CAR index \cite{cannistraci2013link}, which accounts for the role of the local community of common neighbors. The CAR index is defined as
\begin{equation}
	S_{xy}^{CAR}=S_{xy}^{CN} \cdot \sum_{z \in \Gamma(x) \cap \Gamma(y)} \frac{|\gamma(z)|}{2},
	\label{CAR_c}
\end{equation}
where $\gamma(z)$ denotes the subset of neighbors of $z$ that are neighboring to both $x$ and $y$, excluding $x$ and $y$ themselves. It is known that $S_{xy}^{CN}$ is equivalent to $M_2(x,y)$, and $\sum_{z \in \Gamma(x) \cap \Gamma(y)} \frac{|\gamma(z)|}{2}$ is the number of links connecting common neighbors of $x$ and $y$, which is equivalent to the number of M12$(\vcenter{\hbox{\includegraphics[width=2.8ex,height=2.8ex]{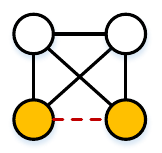}}})$ with $(x,y)$ being the target link (Fig. \ref{Fig2_Mechanisms}c). Therefore, the CAR can be rewritten as
\begin{equation}
	S_{xy}^{CAR} = M_2(x,y) \cdot M_{12}(x,y).
	\label{CAR}
\end{equation}

The CN, AA, and RA indices are first-order similarity indices. Beyond first-order similarity (also called 2-hop-based) algorithms, there are second-order similarity (also called 3-hop-based) indices that incorporate information from 3-hop paths \cite{zhou2009predicting,lu2009similarity,kovacs2019network,zhou2021experimental}. These indices can be expressed using second-order edge orbit degrees.
Herein, we use CN-L3 index \cite{kovacs2019network,zhou2021experimental} as an example, which counts the number of 3-hop paths between the target node pair, as
\begin{equation}
	\begin{split}
		S_{xy}^{CN - L{\rm{3}}}{\rm{ = }}\sum\limits_{i \in \Gamma (x),j \in \Gamma (y)} {{A_{ij}}},
	\end{split}
	\label{CN_3}
\end{equation}
where $A$ is the adjacency matrix. If $i$ and $j$ are directly connected, $A_{ij}=1$; otherwise, $A_{ij}=0$. As illustrated in Fig. \ref{Fig2_Mechanisms}d, the CN-L3 index can be interpreted as a linear combination of the edge orbit degrees M9$(\vcenter{\hbox{\includegraphics[width=2.8ex,height=2.8ex]{./motif/M9.pdf}}})$, M11$(\vcenter{\hbox{\includegraphics[width=2.8ex,height=2.8ex]{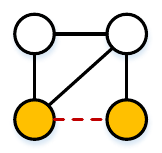}}})$, and M12$(\vcenter{\hbox{\includegraphics[width=2.8ex,height=2.8ex]{./motif/M12.pdf}}})$. In M9$(\vcenter{\hbox{\includegraphics[width=2.8ex,height=2.8ex]{./motif/M9.pdf}}})$ or M11$(\vcenter{\hbox{\includegraphics[width=2.8ex,height=2.8ex]{./motif/M11.pdf}}})$, one 3-hop path connects nodes $x$ and $y$, whereas in M12$(\vcenter{\hbox{\includegraphics[width=2.8ex,height=2.8ex]{./motif/M12.pdf}}})$, two such paths exist, thus the CN-L3 index can be rewritten in terms of second-order edge orbit degrees as
\begin{equation}
	S_{xy}^{CN - L{\rm{3}}} =M_9(x,y)+M_{11}(x,y)+2M_{12}(x,y).
	\label{CN_L3}
\end{equation}

The recently proposed motif-based similarity (MS) \cite{li2021research} accounts for triangular motifs associated with the target pair of nodes, as
\begin{equation}
	S_{xy}^{MS} = \frac{{\left| {\Gamma (x) \cap \Gamma (y)} \right|}}{{\left| {\Gamma (x) \cap \Gamma (y)} \right| + motif(x) + motif(y)}},
\end{equation}
where $motif(x)$ represents the number of triangular motifs that node $x$ partakes in (i.e., a popularity of node $x$). Obviously, such number is exactly the same to the node orbit degree N4$(\vcenter{\hbox{\includegraphics[width=3.3ex,height=3.3ex]{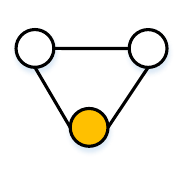}}})$ of $x$ (Fig. \ref{Fig2_Mechanisms}e). Therefore, MS can be expressed by orbit degrees as
\begin{equation}
	S_{xy}^{MS} = \frac{M_2(x,y)}{{M_2(x,y)+ N_4(x) + N_4(y)}}.
	\label{MS}
\end{equation}

In summary, those well-known multi-order popularity-based and similarity-based indices can be represented by orbit degrees, while some structural characteristics that can be captured by orbit degrees have not been formalized by any popularity or similarity indices in the literature. Therefore, we believe that treating orbit degrees as features, combined with some suitable machine learning methods, will yield more accurate predictions. Moreover, by analyzing the importance of orbit-degree-based features in a machine learning model, we can comprehensively uncover the roles of multi-order popularity and similarity mechanisms represented by these features in link prediction.

{\bf Performance Evaluation.}
Our experiments focus on undirected and unweighted networks and cover 550 real-world networks from disparate domains (see \hyperref[sec:Data description]{Materials and Methods}). A network is formally denoted by $G(V,E)$, where $V$ and $E$ represent the sets of nodes and links, respectively. In each implementation of a network $G$, its link set $E$ is randomly divided into three parts---$E=E^{Tr} \cup E^{Va} \cup E^{Te}$---at a ratio of 8:1:1 for training, validation, and testing, respectively. The training set $E^{Tr}$ and validation set $E^{Va}$ are known information, whereas the test set $E^{Te}$ is treated as unknown data, whose information can only be used to evaluate the algorithms' performance. Parametric models are trained on $E^{Tr}$, and the model parameters are determined through link prediction for $E^{Va}$. Subsequently, we combine the training and validation sets (i.e., $E^{Tr} \cup E^{Va}$) and use the parametric models with estimated parameters to predict links in $E^{Te}$. For parameter-free models (such as CN and RA), we directly use information in $E^{Tr} \cup E^{Va}$ to calculate the similarity scores and then predict links in $E^{Te}$. The aforementioned procedure is a standard process in machine learning and provides more reliable results than do traditional methods (see, for example, the procedure described in \cite{lu2011link}) in which $E^{Te}$ is used to determine parameters, which results in the information in $E^{Te}$ no longer being unknown for the algorithm. To mitigate the biases introduced by the highly imbalanced ratio in training models and evaluating algorithms, we employ the negative sampling method \cite{goldberg2014word2vec,yang2020understanding}. Specifically, we regard the sets $E^{Va}$ and $E^{Te}$ as positive samples and randomly select two subsets from the set $U-E$ ($U$ denotes the universal set containing all $|V||V-1|/2$ potential links) to serve as negative samples in each run, and these subsets have sizes equal to those of $E^{Va}$ and $E^{Te}$. Consequently, in calibrating model parameters and assessing algorithm performance, the present link prediction task is transformed into a binary classification problem with an equal number of positive and negative samples.

We propose an orbit-degree-based algorithm (called OD algorithm for short) using XGBoost (which is a well-performed and well-known supervised learning model for binary classification \cite{chen2016xgboost,lei2022forecasting}) to fuse all node and edge orbit degrees, where each orbit degree is treated as a one-dimensional input feature. The results obtained using only node orbit degrees and only edge orbit degrees in XGBoost are presented in \textcolor{blue}{Supplementary Note 2}, exhibiting a slightly poorer performance. The benchmark algorithms for comparison comprise nine representative structure-based and three popular embedding-based methods. The structure-based methods include a popularity-based index PA \cite{barabasi1999emergence}, five local similarity indices CN \cite{liben2007link}, RA \cite{zhou2009predicting}, CN-L3 \cite{zhou2021experimental}, RA-L3 \cite{zhou2021experimental}, and CAR \cite{cannistraci2013link}, the global similarity index Katz \cite{katz1953new}, the motif-based index MS \cite{li2021research} and the Node-GDV index \cite{milenkovic2008uncovering}. To be fair, for each structure-based method, the similarity index is also treated as one input feature for XGBoost. The three embedding-based methods include DeepWalk \cite{perozzi2014deepwalk}, Node2vec \cite{grover2016node2vec} and GraphWave \cite{donnat2018learning}.

Table \ref{Table3_all_results} presents the comparison between the OD algorithm and benchmark methods for the 550 real-world networks, with the top-performing method highlighted in bold. The performance is evaluated by four widely-used metrics for classification, namely the area under the ROC curve (AUC), Precision, Recall, and F1-score (see \hyperref[sec:Evaluation metrics]{Materials and Methods}). Overall, OD stands out among the benchmark algorithms, reflected by the fact that it performs the best under all evaluation metrics.
The structure-based methods except Katz consider only a subset of the features in OD, and thus their performance is naturally inferior to OD. Katz accounts for all paths and achieves the highest performance among all benchmarks in AUC, Recall, and F1-score, but it also performs significantly worse than OD. Specifically, compared with Katz, OD shows an 11.4\% increase in AUC, a 4.9\% increase in Recall, and an 8.7\% increase in F1-score.
This illustrates that there is no need to laboriously mine global information, and instead, to leverage local information by properly fusing node and edge orbit degrees can get remarkably better performance. The RA-L3 is the best among benchmarks under Precision, and OD beats it by 4.8\%. In addition, OD also outperforms these embedding-based methods, demonstrating significant improvements. Within these methods, Node2vec performs better than DeepWalk, likely due to its flexible neighborhood sampling strategy. In contrast, GraphWave lags behind, potentially due to its incompatibility with the link prediction task \cite{cao2019network}.

\begin{table}[htbp]
	\setlength{\abovecaptionskip}{0.cm}
	\setlength{\belowcaptionskip}{-0.cm}
	\renewcommand\tabcolsep{2.0 pt}
	\renewcommand{\arraystretch}{1.0}
	\caption{Performance (mean $\pm$ standard deviation) of the proposed model and some selected benchmark models in terms of AUC, Precision, Recall, and F1-score. The presented results are average values obtained for 550 real-world networks, with 10 independent runs implemented for each network.}
	\centering
	\begin{tabular}{l c c c c c}
		\hline
		\hline
		\multicolumn{2}{c}{Methods}  & {AUC} &  Precision & Recall & F1-score \\
		\hline
		\multirow{8}{*}{Structure-based} & PA  \cite{barabasi1999emergence}  & 0.696 $\pm$ 0.106 & 0.649 $\pm$ 0.119 & 0.668 $\pm$ 0.171 & 0.650  $\pm$ 0.125 \\
		&CN  \cite{liben2007link}   & 0.675 $\pm$ 0.194 & 0.622 $\pm$ 0.358 & 0.603 $\pm$ 0.412 & 0.554 $\pm$ 0.346 \\
		&RA  \cite{zhou2009predicting} & 0.672 $\pm$ 0.193 & 0.630  $\pm$ 0.363 & 0.595 $\pm$ 0.412 & 0.550  $\pm$ 0.346 \\
		&CN-L3  \cite{zhou2021experimental}  & 0.707 $\pm$ 0.182 & 0.749 $\pm$ 0.290  & 0.561 $\pm$ 0.369 & 0.566 $\pm$ 0.320 \\
		&RA-L3  \cite{zhou2021experimental} & 0.711 $\pm$ 0.183 & 0.762 $\pm$ 0.288 & 0.561 $\pm$ 0.367 & 0.570  $\pm$ 0.322 \\
		&CAR \cite{cannistraci2013link} & 0.667 $\pm$ 0.190  & 0.618 $\pm$ 0.360  & 0.593 $\pm$ 0.412 & 0.545 $\pm$ 0.345 \\
		&Katz  \cite{katz1953new}  & 0.766 $\pm$ 0.148 & 0.713 $\pm$ 0.166 & 0.767 $\pm$ 0.167 & 0.734 $\pm$ 0.155 \\
		&MS  \cite{li2021research}  & 0.669 $\pm$ 0.191 & 0.625 $\pm$ 0.362 & 0.592 $\pm$ 0.413 & 0.546 $\pm$ 0.345 \\
		&Node-GDV  \cite{milenkovic2008uncovering} & 0.706  $\pm$ 0.101 & 0.657 $\pm$ 0.104 & 0.658 $\pm$ 0.129 & 0.652 $\pm$ 0.099 \\
		\hline
		\multirow{3}{*}{Embedding-based} & DeepWalk  \cite{perozzi2014deepwalk}  & 0.707 $\pm$ 0.108 & 0.648 $\pm$ 0.136 & 0.624 $\pm$ 0.135 & 0.610 $\pm$ 0.132 \\
		&Node2vec  \cite{grover2016node2vec} & 0.727 $\pm$ 0.172 & 0.683 $\pm$ 0.182 & 0.693 $\pm$ 0.193 & 0.683 $\pm$ 0.182 \\
		&GraphWave  \cite{donnat2018learning}  & 0.536 $\pm$ 0.078 & 0.515 $\pm$ 0.096 & 0.528 $\pm$ 0.143 & 0.515 $\pm$ 0.102 \\
		\hline
		Proposed &OD & \textbf{0.854 $\pm$ 0.130}  & \textbf{0.799 $\pm$ 0.150} & \textbf{0.805 $\pm$ 0.152} & \textbf{0.798 $\pm$ 0.143} \\
		\hline
		\hline
	\end{tabular}
	\label{Table3_all_results}
\end{table}

\begin{figure*}[htbp]
	\setlength{\abovecaptionskip}{0.cm}
	\setlength{\belowcaptionskip}{-0.cm}
	\centering
	\includegraphics[width=0.9\textwidth]{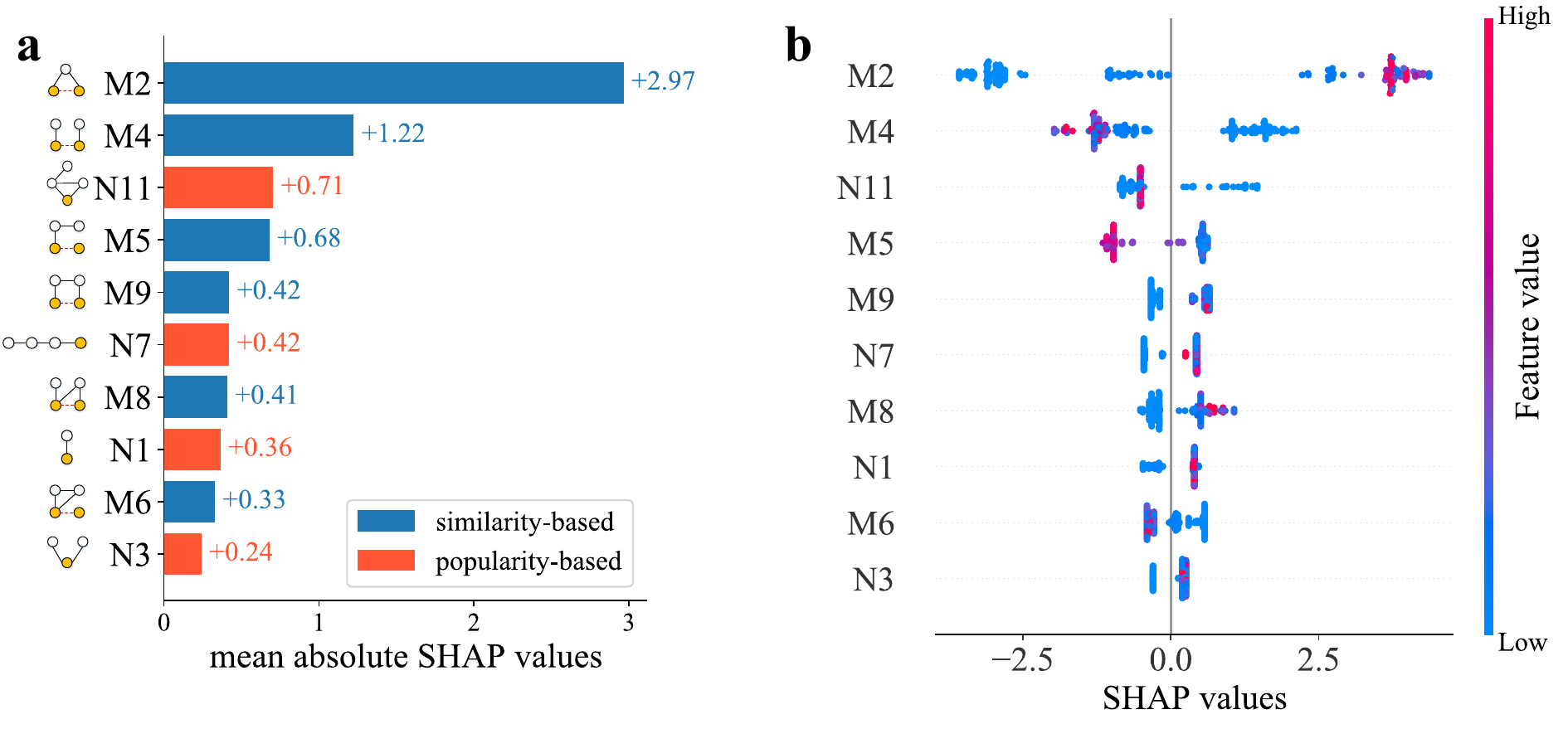}
	\caption{\textbf{Results of feature analyses for the board membership network of Norwegian public limited companies.} \textbf{a} Top 10 features for the XGBoost model, determined on the basis of mean absolute SHAP values. The blue and orange colors denote similarity and popularity features, respectively. The two highest-ranked features, namely M2 and M4, are first-order similarity features, which preliminarily indicates the dominant role of first-order similarity mechanisms in link formation of the examined network. \textbf{b} SHAP values of the top 10 features for all positive and negative samples, with the color of data points reflecting the feature values of corresponding samples.}
	\label{Fig_3_Feature}
\end{figure*}

{\bf Feature analyses.}
To thoroughly analyse the roles of multi-order popularity and similarity mechanisms, we employ Shapley additive explanations (SHAP) values
(see \hyperref[sec:SHAP values]{Materials and Methods}) to quantify the importance of each feature. A higher mean absolute SHAP value of a feature suggests a more important role of that feature, and a positive or negative mean SHAP value of a feature indicates that as the value of this feature increases, the model is more inclined to predict the corresponding sample as a positive or negative instance. Figure \ref{Fig_3_Feature}a displays the top 10 features based on mean absolute SHAP values for the board membership network of Norwegian public limited companies \cite{seierstad2011few}, where two directors are linked if they share membership in at least one board of directors. The two highest-ranked features, specifically M2$(\vcenter{\hbox{\includegraphics[width=2.8ex,height=2.8ex]{./motif/M2.pdf}}})$ and M4$(\vcenter{\hbox{\includegraphics[width=2.8ex,height=2.8ex]{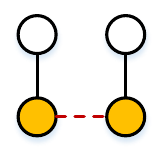}}})$, are identified to be first-order similarity features, with their mean absolute SHAP values are considerably larger than that of the third ranked feature.

Figure \ref{Fig_3_Feature}b depicts the SHAP values of the top 10 features for all positive and negative samples. In this figure, the color of a data point indicates the magnitude of the feature value of the sample, where the color transitions from blue to red as the feature value increases from low to high. If the SHAP value tends to be positive for a higher feature value (indicated by red in Fig. \ref{Fig_3_Feature}b), then the corresponding orbit degree plays a positive role to the existence of a missing link, while if the SHAP value tends to be negative for higher feature values, the corresponding orbit degree can be considered to play a negative role to the existence of a missing link. For the social network analysed in this subsection, an increase in the value of $M_2(x,y)$ increases the probability that $(x,y)$ is a missing link, and in contrast, an increase in the value of $M_4(x,y)$ decreases the probability that $(x,y)$ is a missing link. The detailed quantitative analysis is provided in \textcolor{blue}{Supplementary Note 3}. In summary, through the above analysis, we can evaluate key orbit degree features in any network using features' SHAP values. The absolute SHAP value represents the importance of the feature, while the sign of the SHAP value indicates the direction of the feature's effect on the predicted outcome.

\begin{figure*}[htbp]
	\setlength{\abovecaptionskip}{0.cm}
	\setlength{\belowcaptionskip}{-0.cm}
	\centering
	\includegraphics[width=0.9\textwidth]{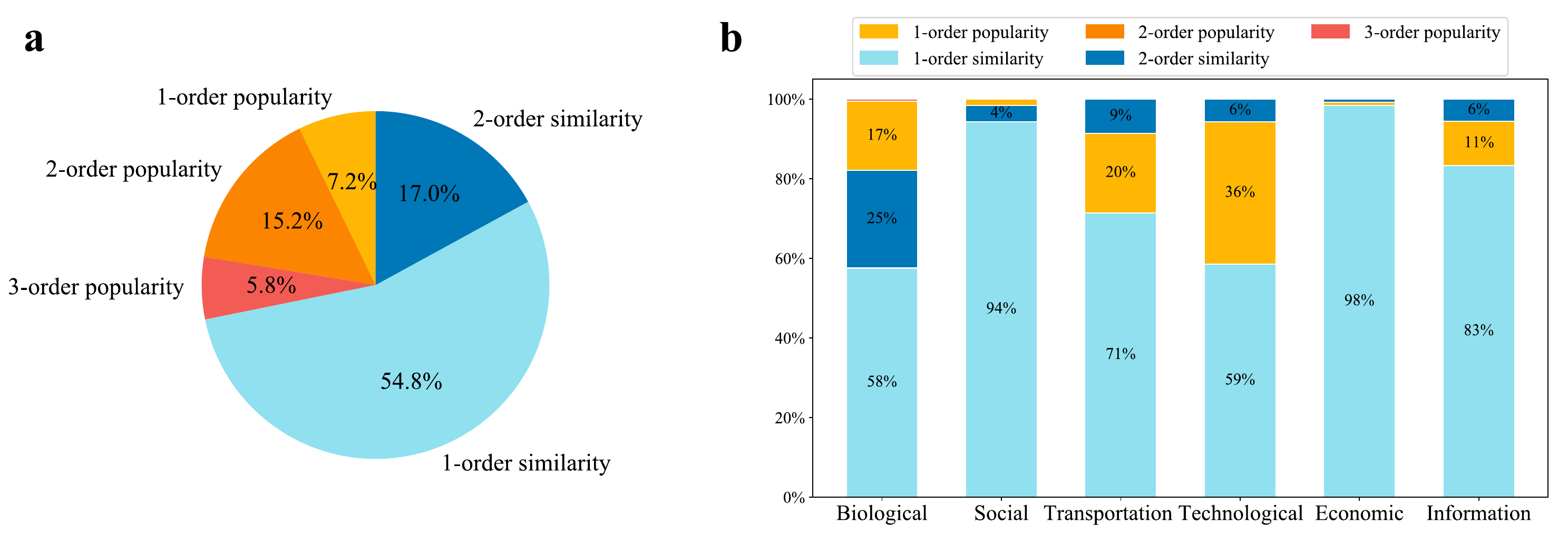}
	\caption{\textbf{Contribution share analysis of multi-order popularity and similarity feature categories.} \textbf{a} The contribution share of five feature categories for the board membership network of Norwegian public limited companies. \textbf{b} The winning rates of the five feature categories in the six domains, which are determined based on their contribution shares in each network.}
	\label{Fig_4_category}
\end{figure*}

{\bf Category analyses.}
From the social network example discussed above, we observe that the mean absolute SHAP values of the first-order similarity features M2$(\vcenter{\hbox{\includegraphics[width=2.8ex,height=2.8ex]{./motif/M2.pdf}}})$ and M4$(\vcenter{\hbox{\includegraphics[width=2.8ex,height=2.8ex]{./motif/M4.pdf}}})$ are significantly larger than those of other features, indicating that first-order similarity plays a primary role in link prediction for this network. To quantify the contribution share of each category, we calculate the ratio of the sum of mean absolute SHAP values for features in that category to the sum of mean absolute SHAP values for all twenty-seven  features (i.e., N1\textendash N15 and M1\textendash M12).  For example, the contribution share of the first-order popularity category is the sum of the mean absolute SHAP values for N1$(\vcenter{\hbox{\includegraphics[width=1.5ex,height=2.8ex]{./motif/N1.pdf}}})$, N3$(\vcenter{\hbox{\includegraphics[width=2.8ex,height=2.8ex]{./motif/N3.pdf}}})$, and N6$(\vcenter{\hbox{\includegraphics[width=2.8ex,height=2.8ex]{./motif/N6.pdf}}})$, divided by the total mean absolute SHAP values of all twenty-seven features. As shown in Fig. \ref{Fig_4_category}a,
we find the first-order similarity features have the highest contribution share, accounting for 54.8\%, which significantly exceeds that of the other four categories. This further underscores the dominant role of first-order similarity features in link prediction for this network, while the other feature categories play more of a complementary role.

We proceed to evaluate the feature categories that contribute the most to each network across different domains. Figure \ref{Fig_4_category}b illustrates the winning rates for five feature categories in each network across six domains. Overall, similarity features show a significant advantage across all domains examined, particularly the first-order similarity, which exhibits consistently high winning rates. Specifically, across the six domains, the winning rates for first-order similarity range from 58\% to 98\%. In economic networks, first-order similarity dominates, contributing the most in 98\% (122 out of 124) of the cases analyzed. Similarly, in social networks, first-order similarity leads with a winning rate of 94\% (117 out of 124). In contrast, popularity features tend to have lower winning rates overall, though they are somewhat more prominent in specific domains. In technological networks, for example, 36\% (24 out of 67) of the analyzed cases show first-order popularity as the leading contributor. On the other hand, for both popularity and similarity features, lower-order features have a more substantial contribution compared to higher-order ones. Notably, only in the biological domain do 25\% (44 out of 179) of networks have second-order similarity as the top contributing feature category, whereas in other domains, the winning rates of second-order similarity features are generally below 10\%. The networks where second-order similarity features are most dominant are primarily the food web and connectome networks of biological domain, with detailed are shown in \textcolor{blue}{Supplementary Note 4}. Meanwhile, higher-order (second- and third-order) popularity features almost never emerge as the top contributors. Therefore, we conclude that lower-order features, especially first-order similarity, are the primary driving factors in network link formation, while higher-order features tend to play a complementary role.

{\bf Domain analyses.}
The results obtained for the board membership network of Norwegian public limited companies indicate that M2$(\vcenter{\hbox{\includegraphics[width=2.8ex,height=2.8ex]{./motif/M2.pdf}}})$ is the most important feature in this network. As M2$(\vcenter{\hbox{\includegraphics[width=2.8ex,height=2.8ex]{./motif/M2.pdf}}})$ has the highest mean absolute SHAP value and positive mean SHAP value, we can infer that this first-order similarity mechanism (also referred to as the homophily mechanism \cite{mcpherson2001birds}) plays a critical role in the link formation of the example network. In a word, by analyzing the dominant feature of a given network, we can gain valuable insights into how the network is structured. Meanwhile, it has long been known that network topological characteristics can vary greatly across different domains \cite{newman2003social,estrada2007topological}, and recent large-scale empirical studies have shown that link prediction algorithms exhibit significant performance disparities when applied to networks from different domains \cite{ghasemian2020stacking,zhou2021experimental,muscoloni2023stealing}. Therefore, we aim to investigate whether similarity and popularity mechanisms play different roles in link prediction across networks from different domains.

\begin{figure*}[htbp]
	\setlength{\abovecaptionskip}{0.cm}
	\setlength{\belowcaptionskip}{-0.cm}
	\centering
	\includegraphics[width=1.0\textwidth]{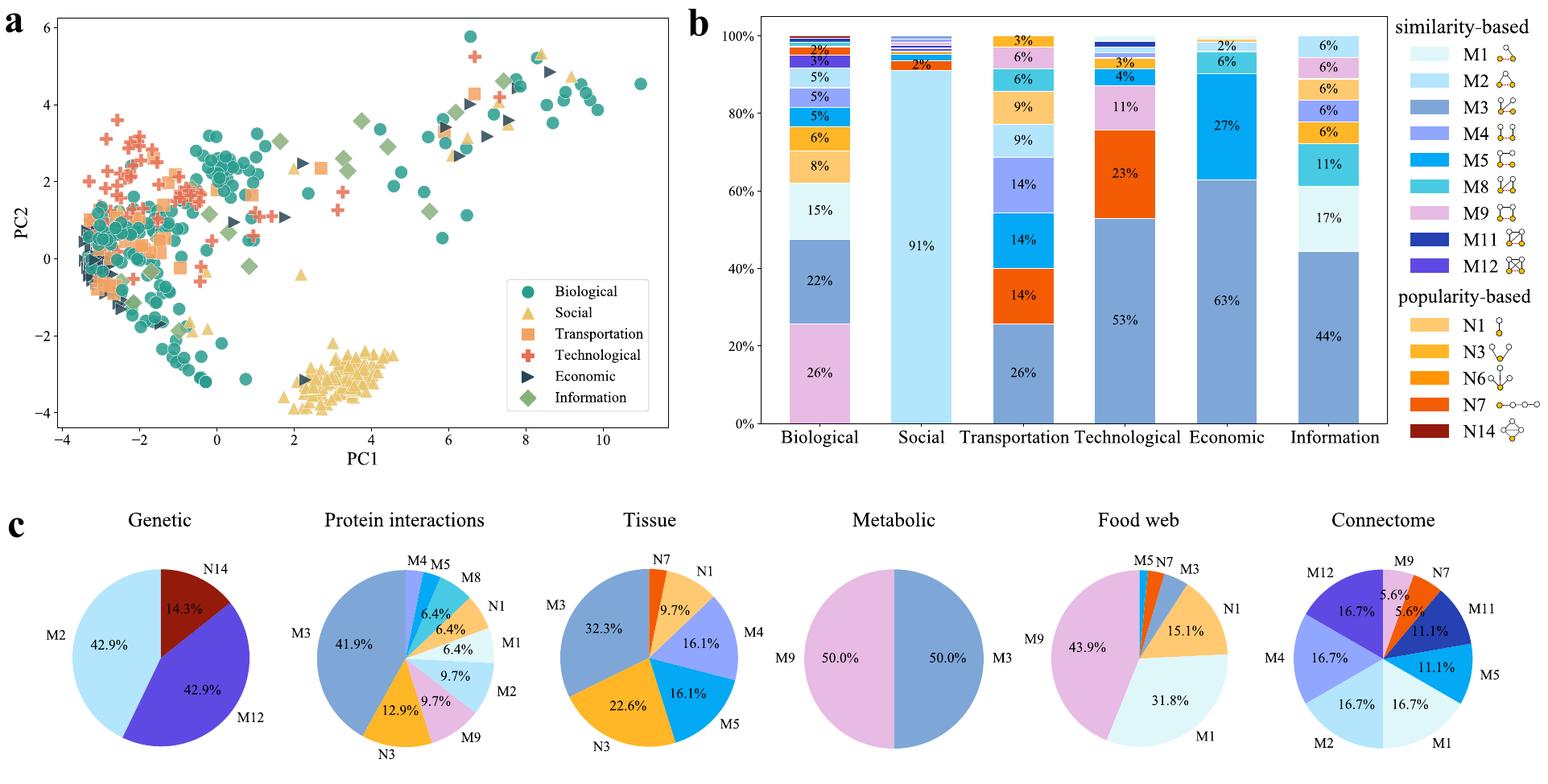}
	\caption{\textbf{Different feature roles in networks from different domains.} \textbf{a} PCA scatter plot of networks. In this scatter plot, each network is represented by a 27-dimensional vector composed of the SHAP values corresponding to the 27 orbit degrees. After these vectors are reduced to two-dimensional ones, data points (corresponding to networks) are visualised with different colors representing different domains.
		\textbf{b} Variations in the winning rates of all features for the six domains, subject to the highest mean absolute SHAP values in each network.
		\textbf{c} The winning rates across all features in biological networks with different sub-domains.
	}
	\label{Fig_5_domain}
\end{figure*}

First, we represent each real network by a 27-dimensional vector in which each element is the SHAP value of the corresponding orbit degree (i.e., N1\textendash N15 and M1\textendash M12). Subsequently, we conduct principal component analysis (PCA) \cite{hotelling1933analysis} to reduce the 550 27-dimensional vectors to two-dimensional ones. As displayed in the PCA scatter plot (Fig. \ref{Fig_5_domain}a), social and economic networks are closely clustered, biological and information networks are dispersed, and transportation and technological networks are moderately clustered. Given a domain, we use the average distance between pairwise networks in the PCA scatter plot to quantify the degree of clustering of networks of this domain, and compare real data and the null model where domain labels are completely randomly shuffled (see \hyperref[sec:Evaluation metrics]{Materials and Methods}). As observed in Table \ref{tab:clustering_degrees}, except for the biological and information networks, networks in any other domain are significantly more clustered than networks in the null model. By directly comparing $\overline{D}$ for different domains, we see social and economic networks are the highest clustered, consistent with the observation in Fig. \ref{Fig_5_domain}a.

\begin{table}[htbp]
	\setlength{\abovecaptionskip}{0.cm}
	\setlength{\belowcaptionskip}{-0.cm}
	\renewcommand\tabcolsep{8.0 pt}
	\renewcommand{\arraystretch}{1.0}
	\centering
	\caption{Degree of clustering of networks from various domains. \(\overline{D}\) represents the average distance in the PCA space over all network pairs within each domain, while \(\langle\tilde{D}\rangle\) denotes the corresponding average distance for the null model over 100 independent runs.}
	\begin{tabular}{l >{\centering\arraybackslash}p{2cm} >{\centering\arraybackslash}p{2cm} >{\centering\arraybackslash}p{2cm}}
		\hline
		\hline
		Domain & \(\overline{D}\) & \(\langle\tilde{D}\rangle\) & $p$-value \\ \hline
		Biological      & 4.215  &  4.537 &0.03\\
		Social          & 1.961   & 4.513  &$<$0.01\\
		Transportation  & 2.568   & 4.461  &$<$0.01\\
		Technological   & 2.386   & 4.579  &$<$0.01\\
		Economic        & 1.882   & 4.497  &$<$0.01\\
		Information   & 4.371   & 4.596  &0.39\\
		\hline
		\hline
	\end{tabular}
	\label{tab:clustering_degrees}
\end{table}

We also examine the first-ranked feature for each network (determined on the basis of the mean absolute SHAP value). Figure \ref{Fig_5_domain}b displays the winning rates of all features in each of the six domains. In 113 of the 124 analysed social networks, M2$\left(\vcenter{\hbox{\includegraphics[width=2.8ex,height=2.8ex]{./motif/M2.pdf}}}\right)$ gets the highest mean absolute SHAP value among all twenty-seven features, thus the winning rate of M2$\left(\vcenter{\hbox{\includegraphics[width=2.8ex,height=2.8ex]{./motif/M2.pdf}}}\right)$ in social networks is $113/124\approx 91\%$. This result demonstrates that the homophily mechanism, a typical first-order similarity mechanism plays a critical role in the link formation of social networks. For the technological, economic, and information networks, the first-order similarity feature M3$\left(\vcenter{\hbox{\includegraphics[width=2.8ex,height=2.8ex]{./motif/M3.pdf}}}\right)$ is the first-ranked feature, with its winning rates being 53\%, 63\%, and 44\%, respectively. In contrast, no single feature holds a significant advantage in biological and transportation networks. As shown in Fig. \ref{Fig_5_domain}c, biological networks from different sub-domains had different first-ranked features, resulting in a broad distribution of dominant features for the biological domain. The case for transportation networks is similar, with detailed sub-domains analysis shown in \textcolor{blue}{Supplementary Note 4}.

\section*{\large\bfseries Discussion}

\hspace*{1em}The proposed graphlet-based predictors provide a rich library for link prediction, where allows existing popularity-based and similarity-based approaches, both lower-order and higher-order, can be considered as specific cases within the present orbit degree framework. Specifically, some lower-order methods (e.g., PA and CN) directly correspond to individual orbit degrees, while higher-order methods (e.g., CN-L3 and CAR) can be equivalently represented by combining a few orbit degrees (Fig. \ref{Fig2_Mechanisms}). By fusing all orbit degree features in the XGBoost model, we remarkably enhance link prediction performance (Table \ref{Table3_all_results}). However, we observe that adding popularity features (node orbit degrees) to similarity features (edge orbit degrees) only marginally improves performance (Table \ref{Table3_all_results} and \textcolor{blue}{Supplementary Table 2}). This suggests that similarity features play a more critical role in link prediction than popularity features, likely because similarity features already encapsulate some aspects of popularity.

We uncover the roles of individual features in link prediction across various domains by calculating features' mean absolute SHAP values in the XGBoost model. Overall, the lower-order features contribute more than higher-order ones in the six domains (Fig. \ref{Fig_4_category}b). First-order similarity consistently emerges as the feature category with the greatest contribution, serving as the primary driving force in link formation across all domains, with winning rates ranging from 58\% in the biological domain to 98\% in the economic domain. Meanwhile, lower-order popularity features also play an important role, achieving notable winning rates in the technological (36\%) and transportation (20\%) domains. In contrast, higher-order features serve a complementary function, with winning rates peaking at just 25\% in the biological domain and remaining below 10\% in the other five domains. Among the few networks where higher-order features contribute most significantly, the primary examples are the food web and connectome networks within the biological domain (\textcolor{blue}{Supplementary Note 4}). On the other hand, among the 124 examined social networks, the homophily mechanism, represented by the first-order similarity feature M2$(\vcenter{\hbox{\includegraphics[width=2.8ex,height=2.8ex]{./motif/M2.pdf}}})$, ranks first in 113 networks, achieving a high winning rate of 91\% (Fig. \ref{Fig_5_domain}b). This demonstrates that homophily mechanism plays a dominant role in the link formation of social networks.
In the domains of economic, technology, and information, M3$(\vcenter{\hbox{\includegraphics[width=2.8ex,height=2.8ex]{./motif/M3.pdf}}})$ feature has the highest winning rate.
M3$(\vcenter{\hbox{\includegraphics[width=2.8ex,height=2.8ex]{./motif/M3.pdf}}})$ is a first-order similarity feature, but it contains some popularity information. As shown in \textcolor{blue}{Supplementary Note 5}, M3$(\vcenter{\hbox{\includegraphics[width=2.8ex,height=2.8ex]{./motif/M3.pdf}}})$ is closely related to but different from PA, and it performs exceptionally well in networks with abundant star-like structures. In contrast, no single feature is significantly prominent in biological and transportation networks. The analysis of the sub-domains of biological and transportation networks reveals significant variations in dominant features (Fig. \ref{Fig_5_domain}c and \textcolor{blue}{Supplementary Note 4}), further underscoring the complexity of link formation mechanisms in these areas.

Our study has two limitations. First, we assess the roles of multi-order popularity and similarity features by simply comparing mean absolute SHAP values rather than conducting an in-depth domain-specific analysis. The roles of some important features, such as M2$(\vcenter{\hbox{\includegraphics[width=2.8ex,height=2.8ex]{./motif/M2.pdf}}})$ and M3$(\vcenter{\hbox{\includegraphics[width=2.8ex,height=2.8ex]{./motif/M3.pdf}}})$, are relatively easily to understand, while for others, we can show whether they are important by SHAP values, but we do not really know their functional roles. Accordingly, future studies should use domain-specific knowledge to conduct in-depth analyses of the roles played by features with high SHAP values for different domains. Second, our study focuses on simple networks, but real systems might be more effectively described by weighted networks \cite{barrat2004architecture}, directed networks \cite{garlaschelli2004patterns}, temporal networks \cite{holme2012temporal}, spatial networks\cite{barthelemy2011spatial}, higher-order networks\cite{bick2023higher}, and so on. Therefore, in future works, we intend to extend the orbit degree framework to the aforementioned more complex networks.

In summary, our work takes into account the structural heterogeneity of nodes and edges within graphlets to achieve high prediction and interpretation performance in link prediction. The applications of graphlets are not limited to link prediction tasks. For instance, we can distinguish the spreading capabilities of different nodes based on the heterogeneity of subgraphs, thereby constructing more precise spreading models \cite{benson2018simplicial,iacopini2019simplicial}. Therefore, in future studies, we hope that the theoretical framework of graphlets will be applied broadly in various issues of network science.

\section*{\large\bfseries Materials and Methods}
\label{sec:Materials and Methods}

\hspace*{1em}{\bf Data description.}
We adopt four standard metrics for binary classification, AUC, Precision, Recall, and F1-score, to evaluate the algorithms' performance. AUC, also called AUROC, AUC-ROC or ROC-AUC, is defined as the area under the Receiver Operating Characteristic (ROC) curve \cite{hanley1982meaning,bradley1997use}, which measures the ability of the model to discriminate between positive and negative classes
across all possible classification thresholds. AUC is the most frequently used metric in link prediction (see, for example, Table 1 in a recent retrospective study \cite{zhou2023discriminating}), probably because of its high interpretability, good visualization, and high discriminating ability \cite{bradley1997use,zhou2023discriminating,jiao2024comparing}. However, some scholars argue that AUC is not suitable for evaluating imbalanced classification problems \cite{yang2015evaluating,saito2015precision}. In the current work, we adopt the negative sampling technique to generate balanced positive and negative classes, so that AUC is a good candidate to evaluate the algorithms' performance. Notice that AUC can be interpreted as the probability that randomly chosen positive samples (missing links) have higher scores than randomly chosen negative samples (nonexistent links), where the scores for the positive and negative samples are the probability of being predicted as a positive sample by the feature-trained XGBoost model. The AUC value should be 0.5 if all scores of the predicted links are randomly generated from an independent and identical distribution, and thus to what extent AUC exceeds 0.5 indicates how much better the algorithm performs than pure chance.

Following the language of confusion matrix in binary classification, model predictions can be classified into four categories:
(1) true positive (TP, missing links predicted to be missing), (2) false positive (FP, nonexistent links predicted to be missing), (3) true negative (TN, nonexistent links predicted to be nonexistent), and (4) false negative (FN, missing links predicted to be nonexistent). Accordingly, Precision \cite{buckland1994relationship} evaluates the accuracy of positive predictions made by a model, calculated as the ratio of true positives to the sum of true positives and false positives, say
\begin{equation}
	\text {Precision}=\frac{TP}{TP+FP}.
\end{equation}
Recall \cite{buckland1994relationship} measures a model's ability to identify all positive samples, defined as the ratio of true positives to the sum of true positives and false negatives, say
\begin{equation}
	\text {Recall}=\frac{TP}{TP+FN}.
\end{equation}
F1-score \cite{sasaki2007truth1} is defined as the harmonic mean of precision and recall, say
\begin{equation}
	F1\text{-}score = 2 \times \frac{\text{Precision} \times \text{Recall}}{\text{Precision} + \text{Recall}}.
\end{equation}

Notice that it is strictly proved that if Precision, Recall, and F1-score adopt the same threshold to predict missing links (i.e., they all treat top-$L$ link with highest scores as missing links), they will give exactly the same ranking of evaluated algorithms \cite{bi2024inconsistency}. However, this work employs a supervised learning framework where each algorithm is tasked with categorizing each sample in the test set as either a positive or negative instance, rather than ranking them based on the probability of being a positive sample. Consequently, different algorithms may not have consistent threshold values for their output results, implying that Precision, Recall, and F1-score could potentially yield differing rankings for these algorithms. In addition, the ranking of algorithms according to AUC is not strongly consistent with the ranking produced by representative threshold-dependent metrics like Precision and Recall (the correlation between AUC and a representative threshold-dependent metric typically ranges from 0.3 to 0.6, measured by the Spearman rank correlation coefficient \cite{bi2024inconsistency}). Therefore, if an algorithm is deemed the best performer across all these metrics, the conclusion about its superiority becomes more robust compared to evaluations based on a single metric alone.

{\bf SHAP values.}
SHAP values, short for SHapley Additive exPlanations, originate from the field of cooperative game theory, which was introduced by the economist Lloyd Shapley in 1953 \cite{shapley1953value}. In the context of machine learning, SHAP values can explain the contribution of each feature to model predictions \cite{lundberg2017unified,meng2021makes}. Thus, SHAP values indicate feature importance. Specifically, for a given sample \( x \), the SHAP value \( \phi_i(x) \) of feature \( i \) is calculated as the weighted average of the contributions of feature \( i \) to model predictions across all possible feature combinations. The condition \( \phi_i(x) > 0 \) implies that feature \( i \) has a positive contribution to the model prediction for sample \( x \), whereas \( \phi_i(x) < 0 \) suggests that feature \( i \) has a negative effect on the model's output. The SHAP value is expressed as follows:
\begin{equation}
	\phi_i(x) = \sum_{S \subseteq F\setminus\{i\}} \frac{|S|!(|F|-|S|-1)!}{|F|!} \left[ f_{S\cup\{i\}}(x_{S\cup\{i\}}) - f_S(x_S) \right],
\end{equation}
where \( F \) denotes the set of all features and \( S \) is a subset of \( F \) that does not include feature \( i \). The function \( f \) represents the prediction model, and \( x_S \) is the value of sample \( x \) when the features in set \( S \) are considered. The term \( f_{S\cup\{i\}}(x_{S\cup\{i\}}) \) represents the model prediction when the features in \( S \) and feature \( i \) are present in the model, whereas \( f_S(x_S) \) represents the model prediction when only the features in \( S \) are present in the model. The SHAP value of feature $i$ is the sum of the SHAP values \( \phi_i(x) \) for all samples $x$.

In our experiments, the SHAP value quantifies the contribution of each feature to the probability of the model predicting a positive sample (missing link), and we use both the positive samples in the test set and the corresponding negative samples to compute the SHAP values of relevant features.

{\bf Clustering in the PCA scatter plot.}
Denoting $d_{ij}$ as the Euclidean distance between two data points $i$ and $j$ in the two-dimensional PCA scatter plot as illustrated in Fig. \ref{Fig_5_domain}a, then the degree of clustering of a set of networks $Q$ can be quantified by the average distance between networks in $Q$, say
\begin{equation}
	\overline{D}_Q= \frac{1}{n(n-1)} \sum_{i,j \in Q, i \neq j} d_{ij}.
	\label{D_domian}
\end{equation}

To investigate the degree of clustering of a given domain of networks $Q$, we can compare $\overline{D}_Q$ with other domains and with its null model. The null model can be obtained by randomly shuffling domain labels. For domain $Q$, the degree of clustering of the null model, which is denoted by $\tilde{D}_Q$, can be obtained by randomly selecting $|Q|$ data points in the PCA plane and then calculating the average distance between these points. If $\tilde{D}_Q$ is significantly larger than $\overline{D}_Q$, we say networks in domain $Q$ are clustered. We perform $S$ independent runs to generate the null model and compare $\overline{D}$ and the average value of $\tilde{D}$ (\(\langle\tilde{D}\rangle\) in Table \ref{tab:clustering_degrees}) for each domain. For any domain $Q$, if the case $\tilde{D}_Q \leq \overline{D}_Q$ occurs for $s$ times, the corresponding $p$-value is $s/S$, and if such case never appears, we write $p<1/S$.

\section*{\large\bfseries Data availability}

\hspace*{1em} The data set used in this study is available in a published article \cite{ghasemian2020stacking}.
The code used in this study is available at \href{https://github.com/Power996/Orbit_degree_for_LP}{https://github.com/Power996/ODLP}.

\section*{\large\bfseries Acknowledgements}

\hspace*{1em} This study was partially supported by the National Natural Science Foundation of China under the grants 62173065 and 42361144718. The high-performance computations conducted in this study were supported by the Interdisciplinary Intelligence Super ComputerDLP. Center of Beijing Normal University, Zhuhai, China.

\section*{\large\bfseries Author contributions}
\hspace*{1em} T.Z. and X.-K.X. conceptualised the research. Y.H. and Y.R. performed the research. Y.H., Y.R., Z.D., T.Z., and X.-K.X. analysed the collected data. Y.H., T.Z., and X.-K.X. wrote the manuscript. Y.R. and Z.D. edited the manuscript.

\section*{\large\bfseries Competing financial interests}
\hspace*{1em} The authors declare no competing financial interests.

\clearpage
\bibliographystyle{unsrt}
\bibliography{Reference1}

\clearpage

\renewcommand{\figurename}{{\bf Supplementary Figure}}
\renewcommand{\thefigure}{{\bf S\arabic{figure}}}
\renewcommand{\thesection}{Supplementary Note \arabic{section}:}
\renewcommand{\thesubsection}{S\arabic{section}\arabic{subsection}}
\renewcommand{\tablename}{{\bf Supplementary Table}}
\renewcommand{\thetable}{{\bf S\arabic{table}}}
\renewcommand{\theequation}{S\arabic{equation}}
\def\msec#1{\bigskip\textbf{#1}}
\def\note#1{{\small\color{red}\textbf{[[#1]]}}}

\makeatletter
\newenvironment{breakablealgorithm}
{
		\begin{center}
			\refstepcounter{algorithm}
			\hrule height.8pt depth0pt \kern2pt
			\renewcommand{\caption}[2][\relax]{
				{\raggedright\textbf{\ALG@name~\thealgorithm} ##2\par}%
				\ifx\relax##1\relax 
				\addcontentsline{loa}{algorithm}{\protect\numberline{\thealgorithm}##2}%
				\else 
				\addcontentsline{loa}{algorithm}{\protect\numberline{\thealgorithm}##1}%
				\fi
				\kern2pt\hrule\kern2pt
			}
		}{
		\kern2pt\hrule\relax
	\end{center}
}

\begin{center}
	\large{\bf Supplementary Information: \\
		Uncovering multi-order popularity and similarity mechanisms in link prediction by graphlet predictors}
\end{center}

\hspace*{\fill}

\hspace*{\fill}

\centerline{Yong-Jian He, Yijun Ran, Zengru Di, Tao Zhou$^*$, and Xiao-Ke Xu$^*$}

\hspace*{\fill}

\centerline{\it $^*$To whom correspondence should be addressed: zhutou@ustc.edu, xuxiaoke@foxmail.com}

\clearpage

\setcounter{figure}{0} 
\setcounter{table}{0} 
\setcounter{equation}{0}

\clearpage
\section{Algorithm implementation and complexity analysis}
\hspace*{1em} Here, we provide a specific implementation of the algorithms to obtain node and edge orbit degrees, along with an analysis of corresponding time complexity.

{\bf Node orbit degrees.}
The calculation of node orbit degrees can be implemented by algorithm \ref{algorithm1}. Assuming that there are $n$ nodes in the network $G$ and the maximum degree of the nodes is $d_{max}$ ($d_{max} \ll n$). First, we need to enumerate all nodes, and then for the $k$-node graphlets, we need to enumerate the neighbors of nodes with $k-1$ times. Therefore, the time complexities of node orbit degrees of 2-node, 3-node, and 4-node graphlets are $O(nd_{max})$, $O(nd_{max}^2)$ and $O(nd_{max}^3)$, respectively.

\renewcommand{\algorithmicrequire}{\textbf{Input:}}
\renewcommand{\algorithmicensure}{\textbf{Output:}}
\begin{breakablealgorithm}
	\caption{Single node orbit degree}
	\begin{algorithmic} [1]                   
		\REQUIRE    $G(V, E)$ : the undirected network \\
		\hspace*{0.50cm} $x$ : the endpoint of predicted edge. \\
		\hspace*{0.50cm}  $Ni$ : the node orbit degree feature \\
		\ENSURE  $r_{a}^{'}$ : the node orbit degree of node $x$ for $Ni$\\
		\vbox{}
		
		\hspace*{-0.6cm} \text {$Nei(a)$} $\leftarrow$ the set of neighbors of any node $a$ in $G$\\
		\hspace*{-0.6cm} Initialize $N1_{num}$ = $N2_{num}$ = $\cdots$  = $N15_{num}$  = 0 \\
		
		\hspace*{-0.6cm} \textbf{For} each node $b$ in $Nei(x)$ \\
		\textbf{If} node $b$ is connected to node $x$ \textbf{do} \\
		\hspace*{-0.6cm} \qquad \qquad $N1_{num}$ += 1 \\

		\textbf{For} each node $c$ in $Nei(b)$ \\
		\qquad \textbf{If} there are links between nodes $b$ and $x$, $c$ and $b$, and no link between $c$ and $x$ \textbf{do} \\
		\qquad\qquad $N2_{num}$ += 1 \\
		
		\qquad\qquad \vdots
		
		\qquad \textbf{Else If} there are links between nodes $b$ and $x$, $c$ and $x$, $b$ and $c$  \textbf{do} \\
		\qquad\qquad $N4_{num}$ += 1 \\
		
		\qquad\textbf{For} each node $d$ in $Nei(c)$ \\
		\qquad\qquad \textbf{If} there are links between nodes $b$ and $x$, $c$ and $b$, $d$ and $b$, and no links between $c$ and $x$, $d$ and $x$, $c$ and $d$  \textbf{do} \\
		\qquad\qquad\qquad $N5_{num}$ += 1 \\
		
		\qquad\qquad \qquad\vdots
		
		\qquad\qquad \textbf{Else If} there are links between nodes $b$ and $x$, $c$ and $x$, $d$ and $x$, $b$ and $c$, $b$ and $d$, $c$ and $d$ \textbf{do} \\
		\qquad\qquad\qquad $N15_{num}$ += 1 \\
		\qquad\textbf{endfor} \\
		\textbf{endfor} \\
		\hspace*{-0.6cm}\textbf{endfor} \\

		\hspace*{-0.6cm}\textbf{If} $Ni$ equals to $N1$  \textbf{do} \\
		$r_{a}^{'}$ = $N1_{num}$ \\
		\qquad\vdots\\
		\hspace*{-0.6cm}\textbf{Else If} $Ni$ equals to $N15$  \textbf{do}\\
		$r_{a}^{'}$ = $N15_{num}$ \\
		\hspace*{-0.6cm}\textbf{return} $r_{a}^{'}$ \\

		\label{algorithm1}
	\end{algorithmic}
\end{breakablealgorithm}

{\bf Edge orbit degrees.}
The calculation of edge orbit degrees can be implemented by Algorithm \ref{algorithm2}. For a given edge $(x,y)$, when calculating the edge orbit degrees of 3-node graphlets, only the neighbors of the node with the greater degree in nodes $x$ and $y$ need to be enumerated. And when calculating the edge orbit degrees of 4-node graphlets, the neighbors of both nodes $x$ and $y$ need to be enumerated. Therefore, the time complexities of edge orbit degrees of 3-node and 4-node graphlets are $O(nd_{max})$ and $O(nd_{max}^2)$, respectively.

\renewcommand{\algorithmicrequire}{\textbf{Input:}}
\renewcommand{\algorithmicensure}{\textbf{Output:}}
\begin{breakablealgorithm}
	\caption{Single edge orbit degree}
	\begin{algorithmic} [1]                   
		\REQUIRE   $G(V, E)$ : the undirected network \\
		\hspace*{0.50cm} $(x,y)$ : the predicted edge\\
		\hspace*{0.50cm}  $Mj$ : the edge orbit degree feature \\
		\ENSURE   $r_{xy}^{'}$ : the edge orbit degree of edge $(x,y)$ for $Mj$ \\
		\vbox{}
		
		\hspace*{-0.6cm}\text {$Nei(a)$} $\leftarrow$ the set of neighbors of any node $a$ in $G$\\
		\hspace*{-0.6cm}\text {$Max\_s(set1,set2)$} $\leftarrow$ the one with more elements in two sets $set1$ and $set2$\\
		\hspace*{-0.6cm}\text {$Min\_s(set1,set2)$} $\leftarrow$ the one with fewer elements in two sets $set1$ and $set2$\\
		
		\hspace*{-0.6cm}Initialize $M1_{num}$ = $M2_{num}$ = $\cdots$ = $M12_{num}$  = 0\\
		\hspace*{-0.6cm}\textbf{For} each node $c$ in $Max\_s(Nei(x),Nei(y))$ \\
		\textbf{If} node $c$ is connected to node $x$ and node $c$ is not connected to node $y$ \textbf{do} \\
		\qquad $M1_{num}$ += 1 \\
		\textbf{Else If} node $c$ is connected to node $x$ and node $c$ is also connected to node $y$ \textbf{do} \\
		\qquad $M2_{num}$ += 1 \\
		
		\textbf{For} each node $d$ in $Min\_s(Nei(x),Nei(y))$ \\
		\qquad \textbf{If} there are links between nodes $c$ and $x$, $d$ and $x$, and no links between $c$ and $y$, $c$ and $d$, $d$ and $y$  \textbf{do} \\
		\qquad\qquad $M3_{num}$ += 1 \\
		
		\qquad\qquad \vdots\\
		\qquad \textbf{Else If} there are links between nodes $c$ and $x$, $d$ and $x$, $c$ and $y$, $c$ and $d$, and $d$ and $y$ \textbf{do} \\
		\qquad\qquad $M12_{num}$ += 1 \\
		\textbf{endfor} \\
		\hspace*{-0.6cm}\textbf{endfor} \\
		
		\hspace*{-0.6cm}\textbf{If} $Mj$ equals to $M1$  \textbf{do} \\
		$r_{xy}^{'}$ = $M1_{num}$ \\
		\qquad\vdots\\
		\hspace*{-0.6cm}\textbf{Else If} $Mj$ equals to $M12$  \textbf{do}\\
		$r_{xy}^{'}$ = $M12_{num}$ \\
		\hspace*{-0.6cm}\textbf{return} $r_{xy}^{'}$ \\

		\label{algorithm2}
	\end{algorithmic}
\end{breakablealgorithm}

\clearpage
\section{Performance based on only node or edge orbit degrees}

\hspace*{1em}In Table 1 of the main text, we show the performance of the XGBoost model using all nodes and edge orbit degrees (N1-N15 and M1-M12). Here we show the performance by using only node orbit degrees (N1-N15) and only edge orbit degrees (M1-M12).

{\bf Performance based on node orbit degrees.}
We carry out link prediction on 550 real-world networks using each individual node orbit degree as well as the fusion of all node orbit degrees (denoted by N1--N15), and compare them with the popularity-based methods, PA \cite{barabasi1999emergence} and Node-GDV \cite{milenkovic2008uncovering}.
The results are presented in Supplementary Table \ref{Table1_node_results}, with the fusion algorithm (i.e., N1--N15, in the last raw) and the best-performed individual degree predictor being emphasized in bold. Among the 15 individual degree predictors, N1 $(\vcenter{\hbox{\includegraphics[width=2.8ex,height=2.8ex]{./motif/N1.pdf}}})$ achieves the highest accuracy (obviously, N1-predictor is same to PA), indicating that the higher-order popularity-based predictors are not superior to lower-order ones. At the same time, the integration of all node orbit degree features yields remarkably better prediction than any individual predictor, illustrating that the fusion of multi-order popularity predictors has a positive effect on the performance of link prediction. Furthermore, such fusion algorithm is also better than Node-GDV, a sophisticated popularity-based algorithm, highlighting the complementary nature of individual features.

\begin{table}[htbp]
	\setlength{\abovecaptionskip}{0.cm}
	\setlength{\belowcaptionskip}{-0.cm}
	\renewcommand\tabcolsep{5.0 pt}
	\renewcommand{\arraystretch}{1.0}
	\caption{Link prediction performance (mean$\pm$SD) of node orbit degrees and existing popularity-based methods, measured by AUC, Precision, Recall, and F1-score. The results are obtained by averaging over the 550 real-world networks, and for each network, we implement 10 independent runs.}
	\centering
	\begin{tabular}{lccccc}
		\hline
		\hline
		Feature & {AUC} &  Precision & Recall & F1-score \\
		\hline
		PA    & 0.696 $\pm$ 0.106 & 0.649 $\pm$ 0.119 & 0.668 $\pm$ 0.171 & 0.650  $\pm$ 0.125 \\
		Node-GDV  & 0.706  $\pm$ 0.101 & 0.657 $\pm$ 0.104 & 0.658 $\pm$ 0.129 & 0.652 $\pm$ 0.099 \\
		\hline
		N1    & \textbf{0.696 $\pm$ 0.106} & \textbf{0.649 $\pm$ 0.119} & \textbf{0.668 $\pm$ 0.171} & \textbf{0.650  $\pm$ 0.125} \\
		N3    & 0.643 $\pm$ 0.093 & 0.602 $\pm$ 0.098 & 0.649 $\pm$ 0.152 & 0.619 $\pm$ 0.108 \\
		N6    & 0.606 $\pm$ 0.098 & 0.571 $\pm$ 0.114 & 0.643 $\pm$ 0.217 & 0.591 $\pm$ 0.142 \\
		\hline
		N2    & 0.602 $\pm$ 0.101 & 0.606 $\pm$ 0.186 & 0.511 $\pm$ 0.292 & 0.504 $\pm$ 0.204 \\
		N4    & 0.556 $\pm$ 0.080 & 0.453 $\pm$ 0.304 & 0.468 $\pm$ 0.383 & 0.407 $\pm$ 0.275 \\
		N5    & 0.565 $\pm$ 0.081 & 0.592 $\pm$ 0.261 & 0.458 $\pm$ 0.375 & 0.423 $\pm$ 0.247 \\
		N8    & 0.587 $\pm$ 0.095 & 0.585 $\pm$ 0.187 & 0.519 $\pm$ 0.297 & 0.500   $\pm$ 0.203 \\
		N10   & 0.546 $\pm$ 0.071 & 0.475 $\pm$ 0.324 & 0.410  $\pm$ 0.389 & 0.368 $\pm$ 0.267 \\
		N11   & 0.553 $\pm$ 0.077 & 0.448 $\pm$ 0.300  & 0.470 $\pm$ 0.385 & 0.405 $\pm$ 0.276 \\
		\hline
		N7    & 0.639 $\pm$ 0.093 & 0.596 $\pm$ 0.094 & 0.675 $\pm$ 0.166 & 0.624 $\pm$ 0.106 \\
		N9    & 0.549 $\pm$ 0.072 & 0.427 $\pm$ 0.255 & 0.491 $\pm$ 0.360  & 0.427 $\pm$ 0.262 \\
		N12   & 0.593 $\pm$ 0.101 & 0.538 $\pm$ 0.305 & 0.485 $\pm$ 0.351 & 0.459 $\pm$ 0.265 \\
		N13   & 0.545 $\pm$ 0.072 & 0.455 $\pm$ 0.325 & 0.422 $\pm$ 0.396 & 0.368 $\pm$ 0.274 \\
		N14   & 0.582 $\pm$ 0.108 & 0.473 $\pm$ 0.323 & 0.484 $\pm$ 0.396 & 0.421 $\pm$ 0.297 \\
		N15   & 0.541 $\pm$ 0.069 & 0.430  $\pm$ 0.321 & 0.442 $\pm$ 0.410  & 0.369 $\pm$ 0.284 \\
		\hline
		\textbf{N1--N15}    & \textbf{0.747 $\pm$ 0.109} & \textbf{0.680  $\pm$ 0.109} & \textbf{0.748 $\pm$ 0.140}  & \textbf{0.707 $\pm$ 0.107} \\
		\hline
		\hline
	\end{tabular}%
	\label{Table1_node_results}%
\end{table}%

{\bf Performance based on edge orbit degrees.}
Likewise, we conduct link prediction on 550 real-world networks using each individual edge orbit degree as well as the fusion of all edge orbit degrees (denoted by M1--M12), and compare them with benchmark methods including CN \cite{liben2007link}, RA \cite{zhou2009predicting}, CN-L3 \cite{zhou2021experimental}, RA\_L3 \cite{zhou2021experimental}, CAR \cite{cannistraci2013link}, MS  \cite{li2021research}, and Katz \cite{katz1953new}. The results are presented in Supplementary Table \ref{Table2_edge_results}, with the fusion algorithm (i.e., M1--M12, in the last raw), the best-performed individual degree predictors, and the best-performed benchmark methods being emphasized in bold. Among the 12 individual predictors, M4$(\vcenter{\hbox{\includegraphics[width=2.8ex,height=2.8ex]{./motif/M4.pdf}}})$ achieves the highest AUC and Recall, while M9$(\vcenter{\hbox{\includegraphics[width=2.8ex,height=2.8ex]{./motif/M9.pdf}}})$ and M1$(\vcenter{\hbox{\includegraphics[width=2.8ex,height=2.8ex]{./motif/M1.pdf}}})$ perform best subject to Precision and F1-score, respectively. Under any evaluation metric, none of the individual predictor outperforms the best benchmark method, while the fusion algorithm always performs better than all benchmarks for all metrics. It again demonstrates the significant improvement by considering multi-order similarity predictors.

\begin{table}[htbp]
	\setlength{\abovecaptionskip}{0.cm}
	\setlength{\belowcaptionskip}{-0.cm}
	\renewcommand\tabcolsep{5.0 pt}
	\renewcommand{\arraystretch}{1.0}
	\caption{Link prediction performance (mean$\pm$SD) of edge orbit degrees and existing similarity-based methods, measured by AUC, Precision, Recall, and  F1-score. The results are obtained by averaging over the 550 real-world networks, and for each network, we implement 10 independent runs.}
	\centering
	\begin{tabular}{lccccc}
		\hline
		\hline
		Feature & {AUC} &  Precision & Recall & F1-score \\
		\hline
		CN & 0.675 $\pm$ 0.194 & 0.622 $\pm$ 0.358 & 0.603 $\pm$ 0.412 & 0.554 $\pm$ 0.346 \\
		RA   & 0.672 $\pm$ 0.193 & 0.630  $\pm$ 0.363 & 0.595 $\pm$ 0.412 & 0.550  $\pm$ 0.346 \\
		RA\_L3  & 0.711 $\pm$ 0.183 & \textbf{0.762 $\pm$ 0.288} & 0.561 $\pm$ 0.367 & 0.570  $\pm$ 0.322 \\
		CN\_L3 & 0.707 $\pm$ 0.182 & 0.749 $\pm$ 0.290  & 0.561 $\pm$ 0.369 & 0.566 $\pm$ 0.320 \\
		CAR  & 0.667 $\pm$ 0.190  & 0.618 $\pm$ 0.360  & 0.593 $\pm$ 0.412 & 0.545 $\pm$ 0.345 \\
		MS  & 0.669 $\pm$ 0.191 & 0.625 $\pm$ 0.362 & 0.592 $\pm$ 0.413 & 0.546 $\pm$ 0.345 \\
		Katz & \textbf{0.766 $\pm$ 0.148} & 0.713 $\pm$ 0.166 & \textbf{0.767 $\pm$ 0.167} & \textbf{0.734 $\pm$ 0.155} \\
		Node-GDV  & 0.706  $\pm$ 0.101 & 0.657 $\pm$ 0.104 & 0.658 $\pm$ 0.129 & 0.652 $\pm$ 0.099 \\
		\hline
		M1    & 0.731 $\pm$ 0.113 & 0.709 $\pm$ 0.142 & 0.678 $\pm$ 0.162 & \textbf{0.681 $\pm$ 0.123} \\
		M2    & 0.675 $\pm$ 0.194 & 0.622 $\pm$ 0.358 & 0.603 $\pm$ 0.412 & 0.554 $\pm$ 0.346 \\
		M3    & 0.722 $\pm$ 0.104 & 0.691 $\pm$ 0.121 & 0.679 $\pm$ 0.163 & 0.675 $\pm$ 0.120 \\
		M4    & \textbf{0.741 $\pm$ 0.130}  & 0.702 $\pm$ 0.149 & \textbf{0.709 $\pm$ 0.160} & 0.696 $\pm$ 0.136 \\
		M8    & 0.650  $\pm$ 0.165 & 0.623 $\pm$ 0.363 & 0.558 $\pm$ 0.389 & 0.525 $\pm$ 0.325 \\
		M10   & 0.595 $\pm$ 0.126 & 0.564 $\pm$ 0.398 & 0.474 $\pm$ 0.404 & 0.433 $\pm$ 0.312 \\
		\hline
		M5    & 0.691 $\pm$ 0.117 & 0.653 $\pm$ 0.130  & 0.634 $\pm$ 0.146 & 0.636 $\pm$ 0.119 \\
		M6    & 0.616 $\pm$ 0.133 & 0.513 $\pm$ 0.289 & 0.559 $\pm$ 0.348 & 0.506 $\pm$ 0.277 \\
		M7    & 0.639 $\pm$ 0.154 & 0.609 $\pm$ 0.358 & 0.549 $\pm$ 0.386 & 0.514 $\pm$ 0.318 \\
		M9    & 0.620  $\pm$ 0.112 & \textbf{0.713 $\pm$ 0.286} & 0.421 $\pm$ 0.316 & 0.450  $\pm$ 0.247 \\
		M11   & 0.636 $\pm$ 0.168 & 0.561 $\pm$ 0.384 & 0.541 $\pm$ 0.405 & 0.496 $\pm$ 0.341 \\
		M12   & 0.619 $\pm$ 0.172 & 0.537 $\pm$ 0.405 & 0.523 $\pm$ 0.431 & 0.464 $\pm$ 0.359 \\
		\hline
		\textbf{M1--M12} & \textbf{0.851 $\pm$ 0.130}  & \textbf{0.796 $\pm$ 0.152} & \textbf{0.802 $\pm$ 0.159} & \textbf{0.794 $\pm$ 0.148} \\
		\hline
		\hline
	\end{tabular}%
	\label{Table2_edge_results}
\end{table}%

{\bf Correlation analysis.}
We use the Pearson correlation coefficient \cite{mudelsee2003estimating} to measure correlations between pairwise orbit degrees. The correlation matrix of the 15 node orbit degrees is shown in the Supplementary Figure. \ref{Fig_Correlation_node}, where node orbit degrees in each of the three black boxes are strongly correlated to each other. For example, 10 of the 14 node orbit degrees are strongly correlated with N1$(\vcenter{\hbox{\includegraphics[width=2.8ex,height=2.8ex]{./motif/N1.pdf}}})$ (see those in the largest black box), and the other 4 are complementary to it (i.e., N3$(\vcenter{\hbox{\includegraphics[width=2.8ex,height=2.8ex]{./motif/N3.pdf}}})$, N6$(\vcenter{\hbox{\includegraphics[width=2.8ex,height=2.8ex]{./motif/N6.pdf}}})$, N9$(\vcenter{\hbox{\includegraphics[width=2.8ex,height=2.8ex]{./motif/N9.pdf}}})$ and N14$(\vcenter{\hbox{\includegraphics[width=2.8ex,height=2.8ex]{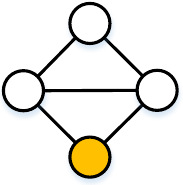}}})$). From such matrix, we can conclude that no single node orbit degree is strongly correlated to all others, so that the fusion of all node orbit degrees provides richer information than any individual degree and thus higher predicting power. Next, we calculate the correlation matrix of the 12 edge orbit degrees based on all node pairs. As shown in Supplementary Figure \ref{Fig_Correlation_edge}, there are three clusters. The first cluster contains M1$(\vcenter{\hbox{\includegraphics[width=2.8ex,height=2.8ex]{./motif/M1.pdf}}})$, M3$(\vcenter{\hbox{\includegraphics[width=2.8ex,height=2.8ex]{./motif/M3.pdf}}})$, M4$(\vcenter{\hbox{\includegraphics[width=2.8ex,height=2.8ex]{./motif/M4.pdf}}})$, M5$(\vcenter{\hbox{\includegraphics[width=2.8ex,height=2.8ex]{./motif/M5.pdf}}})$, and M6$(\vcenter{\hbox{\includegraphics[width=2.8ex,height=2.8ex]{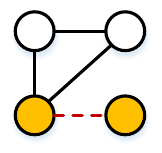}}})$, where the target link is never in a closed triangle or quadrilateral. The second cluster includes M2$(\vcenter{\hbox{\includegraphics[width=2.8ex,height=2.8ex]{./motif/M2.pdf}}})$, M9$(\vcenter{\hbox{\includegraphics[width=2.8ex,height=2.8ex]{./motif/M9.pdf}}})$, M10$(\vcenter{\hbox{\includegraphics[width=2.8ex,height=2.8ex]{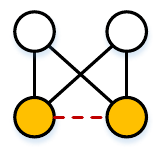}}})$, M11$(\vcenter{\hbox{\includegraphics[width=2.8ex,height=2.8ex]{./motif/M11.pdf}}})$, and M12$(\vcenter{\hbox{\includegraphics[width=2.8ex,height=2.8ex]{./motif/M12.pdf}}})$, where there are 2-hop and/or 3-hop paths connecting the two endpoints of the target link. The last cluster encompasses only two degrees, say M7$(\vcenter{\hbox{\includegraphics[width=2.8ex,height=2.8ex]{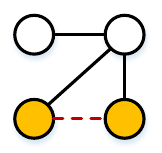}}})$ and M8$(\vcenter{\hbox{\includegraphics[width=2.8ex,height=2.8ex]{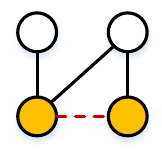}}})$. Analogous to the case of node orbit degrees, the three clusters contain complementary information to each other, therefore the fusion algorithm outperforms those algorithms using one or only a few edge orbit degrees.

\begin{figure}[htbp]
	\setlength{\abovecaptionskip}{0.cm}
	\setlength{\belowcaptionskip}{-0.cm}
	\centering
	\includegraphics[width=0.68\textwidth]{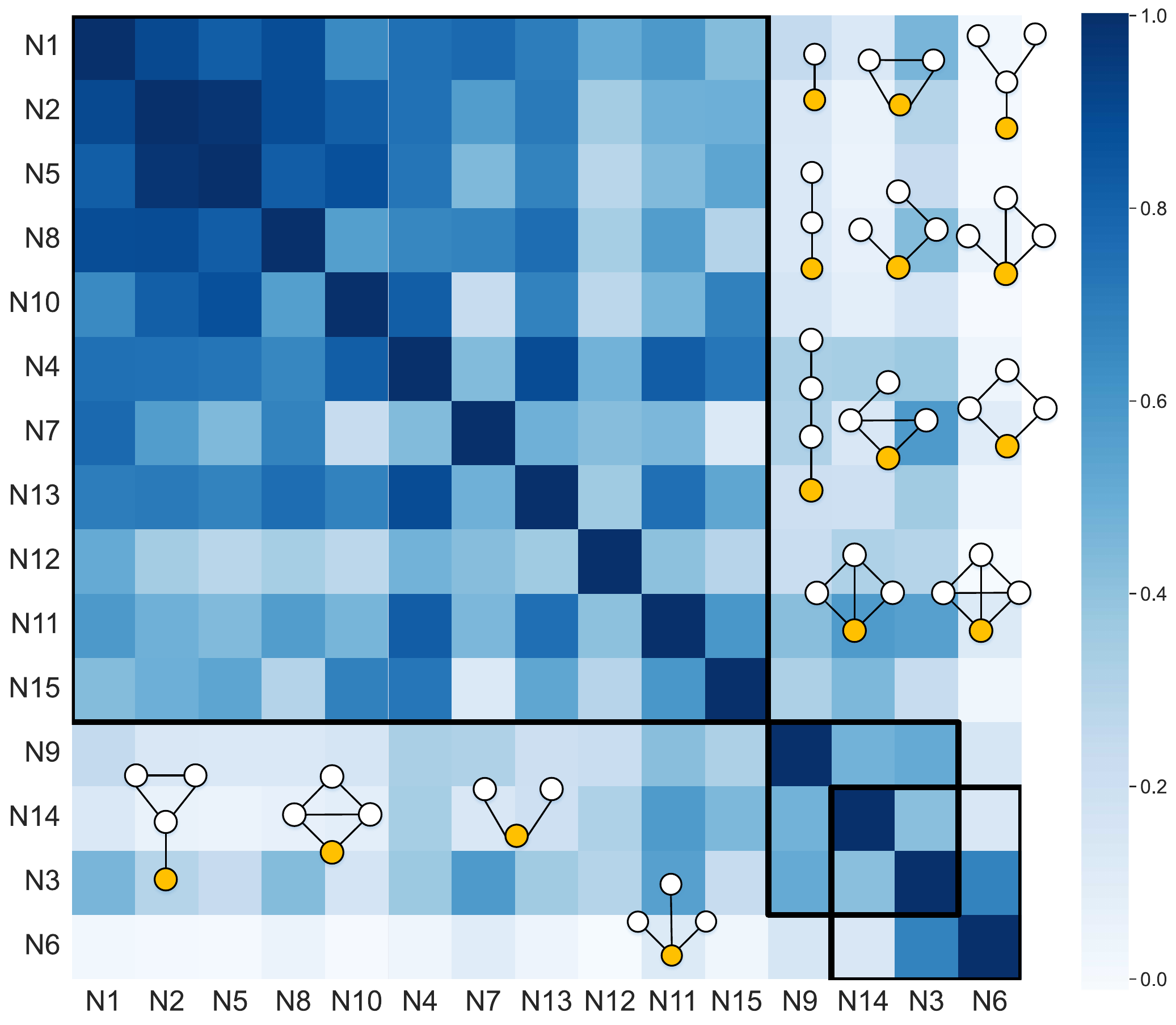}
	\caption{\textbf{The correlation matrix of 15 node orbit degrees for a contact network of high school students.} Although N1, equivalent to PA, is strongly correlated with many node orbit degrees, it has relatively lower correlations to N3, N6, N9, and N14, indicating that some higher-order information, like those contained by N9 and N14, cannot be captured by the popular feature N1, or any other individual 1-order feature.}
	\label{Fig_Correlation_node}
\end{figure}

\begin{figure}[htbp]
	\setlength{\abovecaptionskip}{0.cm}
	\setlength{\belowcaptionskip}{-0.cm}
	\centering
	\includegraphics[width=0.66\textwidth]{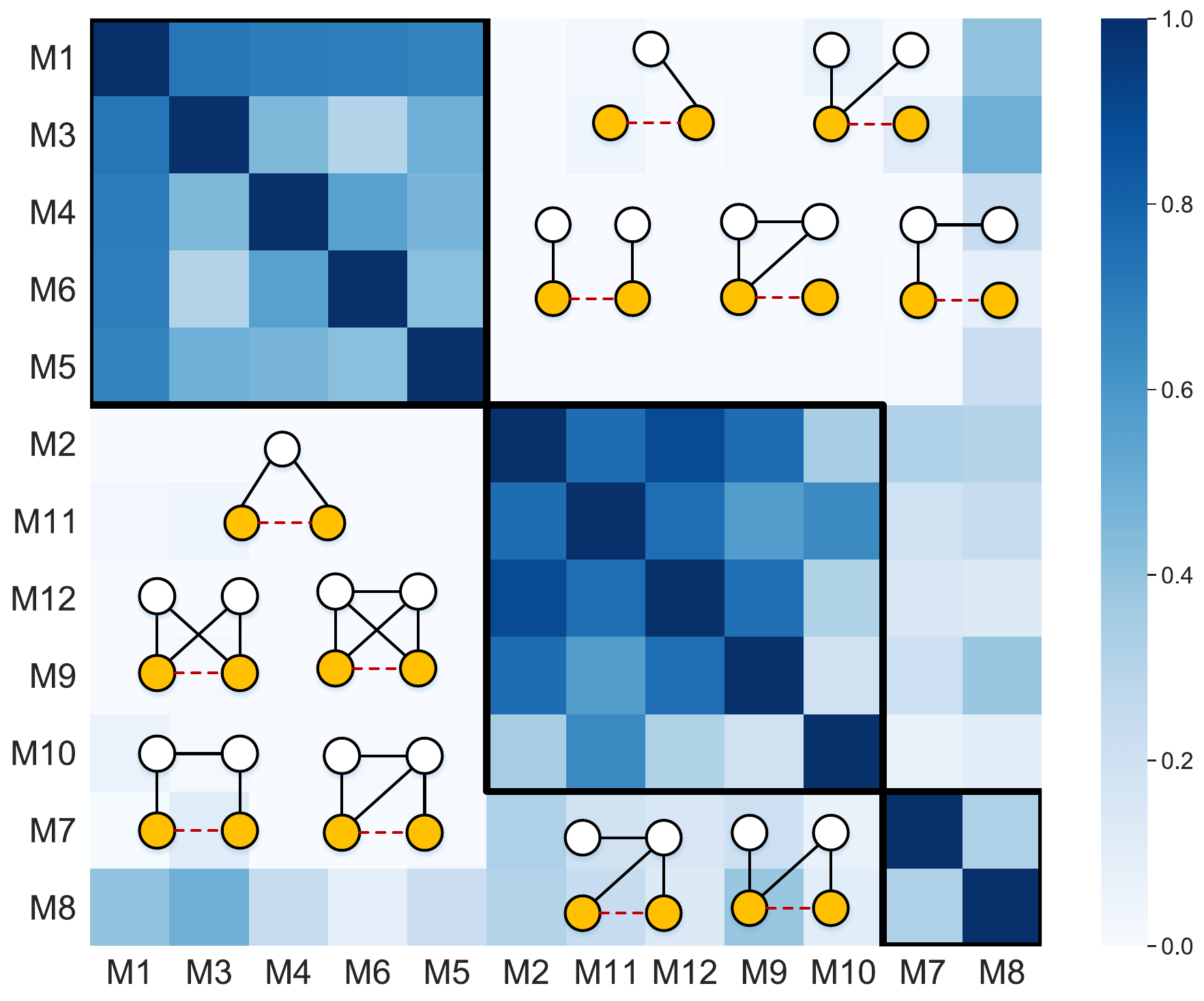}
	\caption{\textbf{The correlation matrix of 12 edge orbit degrees for a contact network of high school students.} The matrix contains three clusters, emphasized by the three black boxes.}
	\label{Fig_Correlation_edge}
\end{figure}

\clearpage
\section{Positive or negative effects of individual features}

\hspace*{1em} In the Fig. 3b of the main text, for a specific social network, we can observe that the feature M2$(\vcenter{\hbox{\includegraphics[width=2.8ex,height=2.8ex]{./motif/M2.pdf}}})$ exhibits a positive effect, say a higher value of $M_2(x,y)$ will increase the probability that $(x,y)$ is a missing link. In contrast, the feature M4$(\vcenter{\hbox{\includegraphics[width=2.8ex,height=2.8ex]{./motif/M4.pdf}}})$ shows a negative effect, namely a higher value of $M_4(x,y)$ will decrease the probability that $(x,y)$ is a missing link. In Supplementary Figure \ref{Feature_values_SHAP}, we further plot the trend between the sample feature values and SHAP values. For M2$(\vcenter{\hbox{\includegraphics[width=2.8ex,height=2.8ex]{./motif/M2.pdf}}})$, the samples with larger feature values are mostly positive samples, and their SHAP values are generally larger than 0, suggesting that the larger the M2$(\vcenter{\hbox{\includegraphics[width=2.8ex,height=2.8ex]{./motif/M2.pdf}}})$ value of a sample, the more likely it is to be a positive sample. This is consistent with the homophily hypothesis in social networks \cite{mcpherson2001birds}, that is, the greater the number of common friends between two nodes, the higher the probability that they are friends. In contrast, for M4$(\vcenter{\hbox{\includegraphics[width=2.8ex,height=2.8ex]{./motif/M4.pdf}}})$, samples with larger feature values are mostly negative samples, and their SHAP values are mostly less than 0. That is to say, the larger the M4$(\vcenter{\hbox{\includegraphics[width=2.8ex,height=2.8ex]{./motif/M4.pdf}}})$ value of a sample, the more likely it is to be a negative sample. Looking closely at $M_4(x,y)$, a larger $M_4(x,y)$ is statistically associated with a larger $k_x$, a larger $k_y$, a fewer common neighbors of $x$ and $y$, and a fewer direct connections between $x$'s neighbors and $y$'s neighbors. As is well known, the formation of most real-world networks adheres to a crucial principle known as the locality principle. Link prediction algorithms based on 2-hop paths (i.e., common neighbors) \cite{liben2007link,zhou2009predicting}, 3-hop paths \cite{kovacs2019network,pech2019link,zhou2021experimental} and local community paradigm \cite{cannistraci2013link}, all rely on this principle. Since the last two factors related to a larger $M_4(x,y)$ are contrary to the locality principle, a larger $M_4(x,y)$ will statistically decrease the likelihood of a link connecting $x$ and $y$.

\begin{figure*}[htbp]
	\setlength{\abovecaptionskip}{0.cm}
	\setlength{\belowcaptionskip}{-0.cm}
	\centering
	\includegraphics[width=0.98\textwidth]{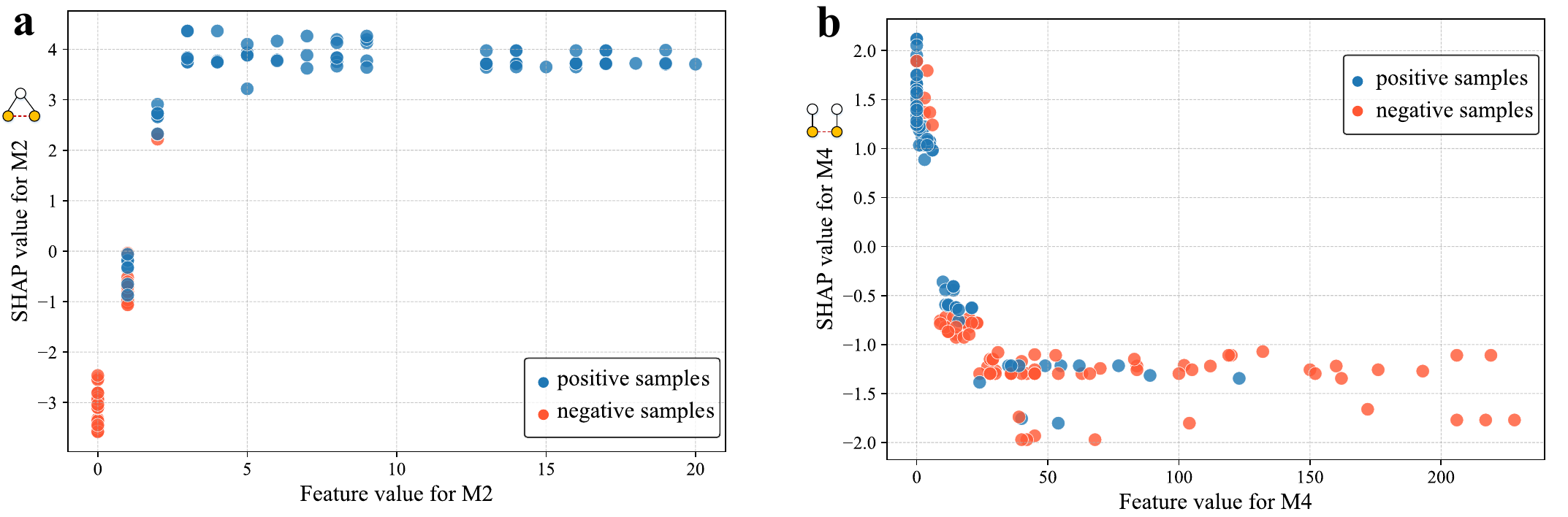}
	\caption{\textbf{The relationship between sample feature values and SHAP values.} Each point represents a sample, with blue indicating positive samples and orange indicating negative samples.}
	\label{Feature_values_SHAP}
\end{figure*}

\clearpage
\section{Analysis of networks dominated by high-order features}

In the Fig. 4b of main text, we observe that low-order features contribute significantly in most networks, while some networks also exhibit substantial contributions from high-order similarity features, such.  For instance, in the biological networks, the winning rate for 2-order similarity features is 25\%. To identify the number of networks where 2-order similarity is the most contributing feature category, we conduct a statistical analysis, with results shown in Supplementary Table \ref{Table3_higher}. The analysis reveals that there are 26 food web networks with a winning rate of 39.4\% in that sub-domain, and 7 connectome networks with a winning rate of 38.9\% in their respective sub-domain. Networks in other domains are fewer in number.

\begin{table}[h]
	\setlength{\abovecaptionskip}{0.cm}
	\setlength{\belowcaptionskip}{-0.cm}
	\renewcommand\tabcolsep{16.0 pt}
	\renewcommand{\arraystretch}{1.0}
	\centering
	\caption{Number of networks where 2-order similarity is the most contributing feature category and their winning rates in respective sub-domains.}
	\begin{tabular}{l c c}
		\hline
		\hline
		Domain/sub-domain & Count & Winning rate in sub-domain \\
		\hline
		Biological/Food web & 26 & 39.4\% (26/66) \\
		Biological/Metabolic & 4 & 15.4\% (4/26) \\
		Biological/Connectome & 7 & 38.9\% (7/18) \\
		Biological/Protein interactions & 4 & 12.9\% (4/31) \\
		Biological/Genetic & 3 & 42.9\% (3/7) \\
		\hline
		
		Economic/Employment &1& 33.3\% (1/2)\\
		\hline
		Transportation/Roads & 2 & 15.4\% (2/13) \\
		Transportation/Public Transport & 1 & 5.0\% (1/20) \\
		\hline
		Technological/Software & 3 & 20.0\% (3/15) \\
		Technological/Digital Circuit & 1 & 2.8\% (1/36) \\
		\hline
		Informational/Language & 1  & 25.0\% (1/4) \\
		\hline
		\hline
		\label{Table3_higher}
	\end{tabular}
\end{table}

\clearpage
\section{Analyzing transportation sub-domain}

\hspace*{1em} Similar to Fig. 5c of the main text, we show the winning rates of features in sub-domains of transportation networks. As shown in Supplementary Figure \ref{subdomain_Transportation}, the first-ranked features in different sub-domains of transportation domains are largely different, illustrating the more complex formation mechanisms of transportation networks than social networks, where the latter are dominated by the homophily mechanism. 
\begin{figure*}[htbp]
	\setlength{\abovecaptionskip}{0.cm}
	\setlength{\belowcaptionskip}{-0.cm}
	\centering
	\includegraphics[width=0.7\textwidth]{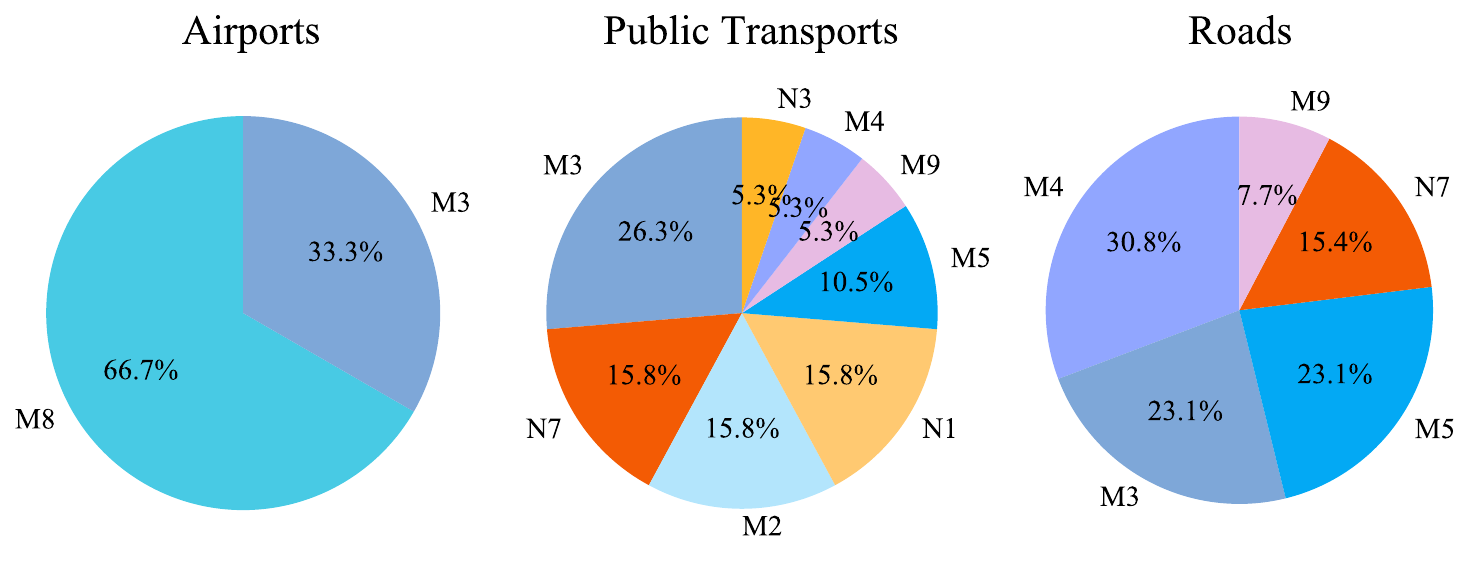}
	\caption{The winning rates subject to the highest mean absolute SHAP values of all features in sub-domains of transportation networks.}
	\label{subdomain_Transportation}
\end{figure*}

\clearpage
\section{M3 and star structure}

\hspace*{1em} In Fig. 5b of the main text, we show that M3$(\vcenter{\hbox{\includegraphics[width=2.8ex,height=2.8ex]{./motif/M3.pdf}}})$ has the highest winning rate across economic, technological, and information networks. Obviously, in M3$(\vcenter{\hbox{\includegraphics[width=2.8ex,height=2.8ex]{./motif/M3.pdf}}})$, the target link can be considered as a link in a star network. To have an intuitive understanding, we visualize an example economic network detailing affiliations among Norwegian public limited companies and their board directors \cite{ghasemian2020stacking}. As shown in Supplementary Figure \ref{net_549_visial}, this network contains many local stars, and any link associated with a hub node in a local star will have a high $M3$ value. Therefore, the M3 predictor performs well in this network.

\begin{figure*}[htbp]
	\setlength{\abovecaptionskip}{0.cm}
	\setlength{\belowcaptionskip}{-0.cm}
	\centering
	\includegraphics[width=0.8\textwidth]{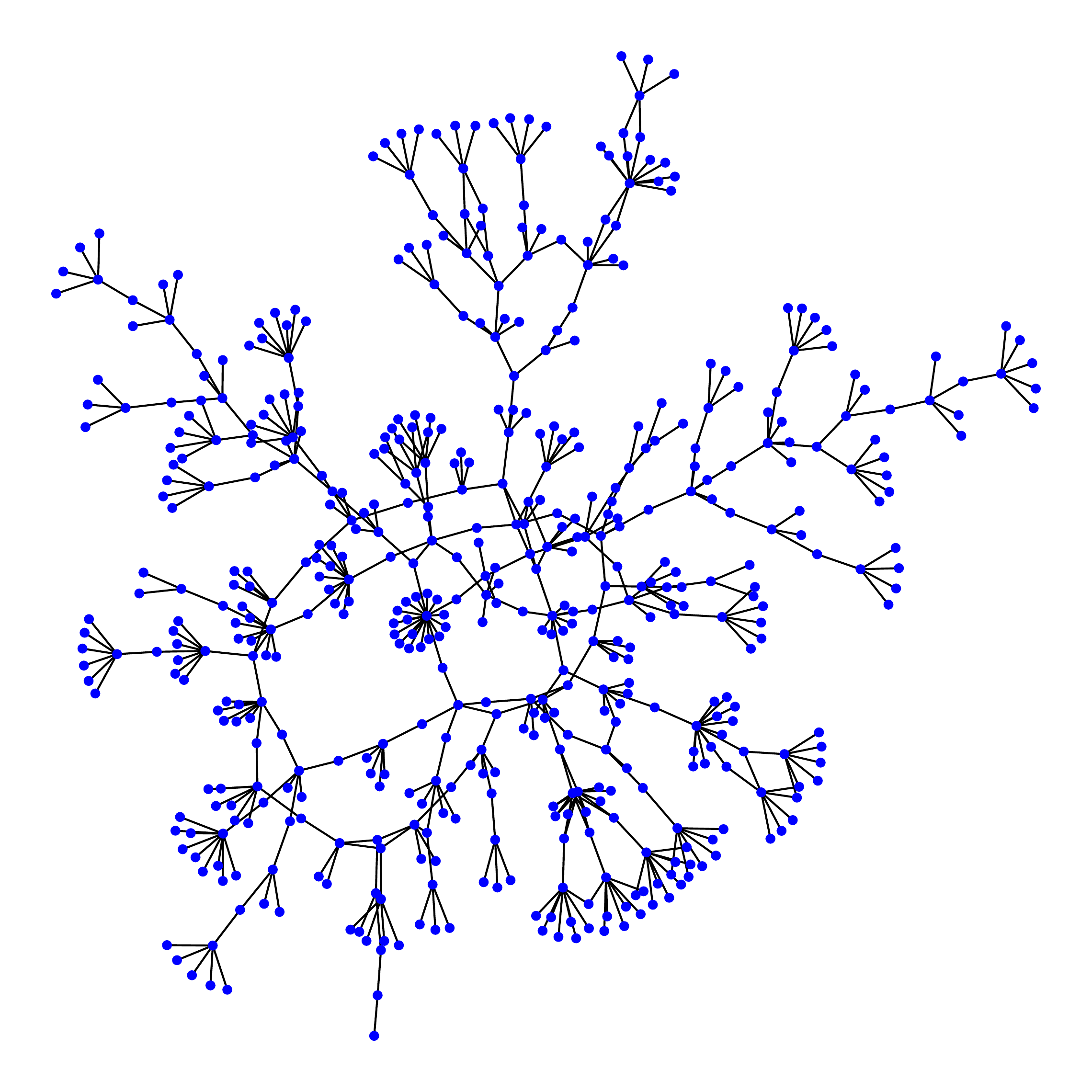}
	\caption{\textbf{Visualization of the example Norwegian economic network.}  This network exhibits rich local star-like structures, which can be well characterized by M3.}
	\label{net_549_visial}
\end{figure*}

\end{document}